\documentclass[aps,prl,twocolumn,groupedaddress]{revtex4-1}
\usepackage[english]{babel}
\usepackage[utf8]{inputenc}
\usepackage[dvips]{graphicx}
\usepackage{amsmath}
\usepackage{setspace}
\usepackage{float}
\usepackage[figuresright]{rotating}
\begin{document}

\title{Electronic band structure of polytypical nanowhiskers: a theoretical approach 
based on group theory and k$\cdot$p method}
\author{Faria Junior, P. E. and Sipahi, G. M.}
\affiliation{Instituto de F\'isica de S\~ao Carlos, USP, S\~ao Carlos, SP, Brazil}
\date{\today}

%================================================================================

\begin{abstract}
Semiconductor nanowhiskers made of III-V compounds exhibit great potential for 
technological applications. Controlling the growth conditions, such as temperature 
and diameter, it is possible to alternate between zinc blend and wurtzite crystalline 
phases, giving origin to the so called polytypism. This effect has great influence in the 
electronic and optical properties of the system, generating new forms of confinement 
to the carriers. A theoretical model capable to accurately describe electronic and 
optical properties in these polytypical nanostructures can be used to study and develop 
new kinds of nanodevices. In this study, we present the development of a wurtzite/zincblend 
polytypical model to calculate the electronic band structure of nanowhiskers based on 
group theory concepts and the k$\cdot$p method. Although the interest is in polytypical 
superlattices, the proposed model was applied to a single quantum well of InP to extract 
the physics of the wurtzite/zincblend polytypism. By the analysis of our results, some trends 
can be predicted: spatial carriers' separation, predominance of perpendicular polarization 
(xy plane) in the luminescence spectra and interband transition blueshifts with strain. A possible 
range of values for the WZ InP spontaneous polarization was suggested.
\end{abstract}

\maketitle

%================================================================================

\section{I. Introduction}

Low dimensional semiconductor structures exhibit different characteristics, ruled 
by the size and morphology of the system. Recently, there is an increasing interest 
in nanowhiskers (NWs), also known as nanowires. These are one dimensional nanostructures 
grown perpendicular to the surface of the substrate, usually by the vapor-liquid-crystal 
(VLC) method. The technological applications of NWs, including biological and chemical 
nanosensors \cite{science-293-1289, nanomedicine-1-51, nbt-23-1294}, lasers 
\cite{nature-421-241}, light emission diodes \cite{nature-409-66} and field effect 
transistors \cite{nl-4-1247}, can be used in a large variety of fields.

The first register in the literature of whiskers was made by Wagner and Ellis \cite{apl-4-89}
 in 1964. In this classic study, it is demonstrated the vertical growth of a Si whisker in 
the [111] direction, activated by droplets of Au, using the VLC method. The radius of 
the structure is approximately the same as the catalist droplet of Au and the vertical 
size depends on the ratio and time of the compound's deposition on the substrate. Although 
the VLC method is the most common, other methods like vapor-phase epitaxy (VPE), 
molecular-beam epitaxy (MBE) and magnetron deposition (MD) are also applied for the NWs growth.

In III-V compound NWs (e. g., arsenides and phosphides), a surprising characteristic is 
the predominance of the wurtzite (WZ) phase in the structure. Exception made for the nitrides, 
the stable crystalline structure of III-V compounds in the bulk form is zincblend (ZB). 
Although the difference between the small formation energy of the two phases are small, 
approximately of $20\;\text{meV}$ per pair of atoms at zero pressure, high pressures would 
be necessary to obtain the WZ phase in the bulk form. However, reducing the dimensions of 
the system to the nanoscale level, such as in these NWs, the WZ phase becomes more stable. 
This stability is due to the smaller surface energy of lateral faces compared to the cubic 
crystal. An extensive summary of NWs growth, properties and applications was made by 
Dubrovskii {\it et al.} \cite{semiconductors-43-1539}.

Controlling the growth conditions, such as temperature and diameter of the NW, it is 
possible to create different regions composed of ZB and WZ structures \cite{nl-10-1699, 
naturemat-5-574, naturenano-4-50, nanoIEEE-6-384, am-21-3654, nano-22-265606, nl-11-2424, 
sst-25-024009}. The mixture of both crystalline phases in the same nanostructure is 
called polytypism. Such characteristic directly affects the electronic and optical 
properties of NWs. The detailed study of polytypism in III-V semiconductor NWs is 
fundamental to the development of novel functional nanodevices with a variety of 
features.

The theoretical tool to calculate the electronic band structure of polytypical NWs 
used in this paper is the k$\cdot$p method. Although the formulation of this method has 
already been done for ZB and WZ crystal structures in the bulk form \cite{pr-100-580, 
book-kane, spjetp-14-898, book-birpikus, prb-54-2491} and in superlattices and heterostructures 
\cite{IEEEjqe-22-1625,prb-53-9930,apl-76-1015,sst-12-252,jcg-246-347}, it was never applied 
to a polytypical case.

Deeply studying the core of the k$\cdot$p formulation for both crystal structures and 
relying on the symmetry connection of the polytypical interface presented in the paper 
of Murayama e Nakayama \cite{prb-49-4710}, it was possible to describe the top valence 
bands and lower conduction bands of ZB and WZ in the same Hamiltonian matrix. The 
envelope function scheme was then applied to obtain the variation of the parameters 
along the growth direction, describing the different regions in a NW, thus completing 
the polytypical model. The effects of strain, spontaneous polarization and piezoelectric 
polarization are also included in the model.

In order to test the model, it was applied to a polytypical WZ/ZB/WZ quantum well of InP. 
Although a real NW is composed by several regions of WZ and ZB, the physical trends of the 
polytypical interface can be extracted from a single quantum well system. We choose the InP 
compound basically for two reasons: the small spin-orbit energy makes easier to fit the matrix 
parameters to the effective mass values given in the paper of De and Pryor \cite{prb-81-155210} 
for the WZ polytype and also because the InP NWs can be found in a great number of studies in 
the present literature \cite{nl-9-648, nano-20-225606, nl-10-1699, prb-82-125327, nanolet-10-4055, 
nanotec-21-505709, ssc-151-781, jap-104-044313}.

The present paper is divided as follows: In section II we discuss the symmetry of ZB 
and WZ crystal structures and analyze how the irreducible representations of the energy 
bands are connected in the polytypical interface. Section III describes the Hamiltonian 
terms for the polytypical model. The results, and their discussion, of InP WZ/ZB/WZ single 
well are found in section IV. Finally, in section V, we draw our conclusions.

%================================================================================

\section{II. Symmetry analysis}

\subsection{A. Zincblend and wurtzite structures}

Our formulation relies on group theory concepts and therefore it is necessary to 
understand the symmetry of the two crystal structures considered in the polytypical 
NWs. The ZB structure belongs to the class of symmorphic space groups and has the 
$T_{d}$ symmetry as its factor group. The number of atoms in the primitive unit 
cell is two. Unlike ZB, the WZ structure belongs to the class of nonsymmorphic 
space groups and its factor group is isomorphic to $C_{6v}$. The classes of symmetry 
operations $C_{2}$, $C_{6}$ and $\sigma_{v}$ are followed by a translation of $c/2$ 
in the [0001] direction. WZ has four atoms in the primitive unit cell. Comparing 
the factor groups one can notice that $C_{6v}$ is less symmetric than $T_{d}$. 
In the k$\cdot$p framework, this lower symmetry decreases the number of irreducible 
representations (IRs) in the group consequently increasing the number of interactions 
in the Hamiltonian. A good description of the concepts of space group symmetry can 
be found in Ref. \cite{dresselhaus-jorio}.

\begin{figure}[h]
\includegraphics{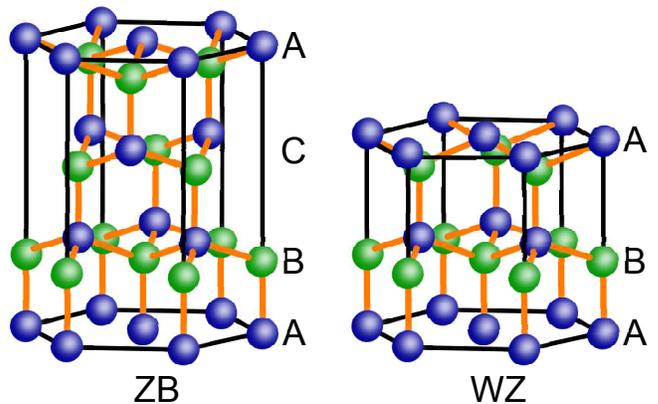}
\caption{ZB (left) and WZ (right) structures and their stacking sequence. The ZB
is presented in the [111] direction. In this direction, the unit cell is twice as 
large as the usual ZB unit cell \cite{prb-49-4710}.}
\label{fig:zb_wz}
\end{figure}

In polytypical NWs, the common growth direction is the ZB [111], which exhibit a 
noticeable similarity to WZ [0001]. Actually, analyzing both crystal structures in these 
directions, one can describe then as stacked hexagonal layers. The ZB has three 
layers in the stacking sequence (ABCABC) while WZ has only two (ABAB) as shown 
in Figure \ref{fig:zb_wz}. The crystal structure alternation occurs when a stacking 
fault happens in WZ, leading to a single ZB segment, or when two twin planes appear 
in ZB, originating a single WZ segment \cite{naturenano-4-50}.

%--------------------------------------------------------------------------------

\subsection{B. Irreducible representations at the polytypical interface}

An important issue of the model is how to connect the energy levels at the 
polytypical interface depending on the symmetry they assume. Based on the scheme 
presented by Murayama and Nakayama \cite{prb-49-4710} of single group IRs at the WZ/ZB 
interface the symmetry of the energy bands can be chose. The same scheme was constructed 
by De and Pryor \cite{prb-81-155210} for the double group IRs with the inclusion of the 
spin-orbit coupling.

\begin{figure}[h]
\includegraphics{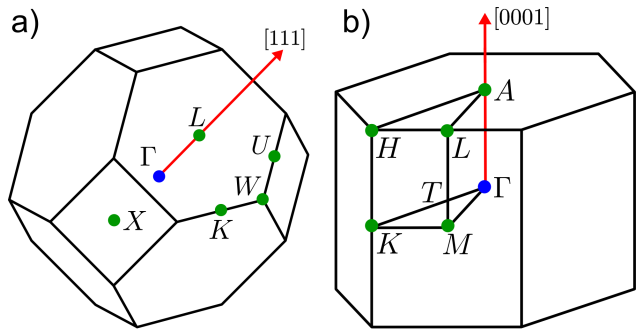}
\caption{Usual a) ZB and (b) WZ first Brillouin zones and their respective high symmetry
points. The arrows (red line) represent the growth directions in NWs. For ZB, the
[111] direction is directed towards the L-point.}
\label{fig:brillouin_zones}
\end{figure}

Since WZ has twice more atoms in the primitive unit cell than ZB, the number of 
energy bands in the $\Gamma$-point is also twice as large. Considering the $sp^3$ 
hybridization, without spin, ZB has 8 energy bands while WZ has 16. However, in the 
[111] direction, the ZB unit cell is two times larger than the usual face-centered 
cubic (FCC) unit cell \cite{prb-49-4710}. In the IRs scheme mentioned above, 
the presence of energy bands with $L$ symmetry takes into account the mismatch in 
the number of atoms for the usual unit cells. The reason for the appearance of the 
$L$ symmetry is the fact that ZB [111] growth direction is directed towards the $L$-point, 
as displayed in Figure \ref{fig:brillouin_zones}a, hence this point is mapped out in 
the $\Gamma$-point. Figure \ref{fig:brillouin_zones}b, displays the first Brillouin zone 
(FBZ) for the WZ structure.

\begin{figure}[h]
\includegraphics{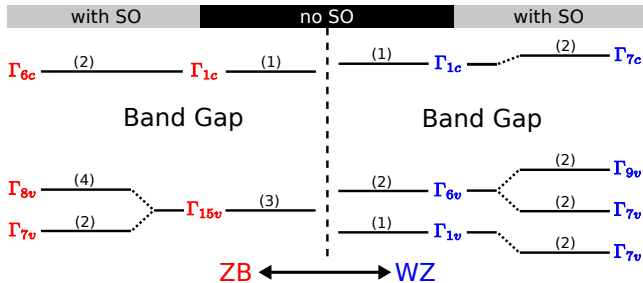}
\caption{The subset of IRs considered in this formulation with and without the
spin-orbit (SO) coupling. The numbers in parentheses are the degeneracy of the IRs.
The notation for the IRs follow Refs. \cite{prb-49-4710,prb-81-155210}}
\label{fig:gamma_symmetry}
\end{figure}

Among all the IRs presented in Refs. \cite{prb-49-4710,prb-81-155210} we 
considered only a small subset. Displayed in Figure \ref{fig:gamma_symmetry}, this subset 
comprises the lower conduction band and the top three valence bands, which belong to the 
$\Gamma$-point in both structures. The price that is paid in considering only a small subset 
is the accuracy of the Hamiltonian for a fraction of the FBZ, approximately 
10-20\%. The basis states for the IRs of the considered bands are presented in equations 
(\ref{eq:ZB_bands_sym}) and (\ref{eq:WZ_bands_sym}) for ZB and WZ, respectively.

\begin{eqnarray}
\Gamma_{1c}^{ZB} & \sim & x^{2}+y^{2}+z^{2}\nonumber \\
\Gamma_{15v}^{ZB} & \sim & \left(x,y,z\right)
\label{eq:ZB_bands_sym}
\end{eqnarray}

\begin{eqnarray}
\Gamma_{1c}^{WZ} & \sim & x^{2}+y^{2}+z^{2}\nonumber \\
\Gamma_{6v}^{WZ} & \sim & \left(x,y\right)\nonumber \\
\Gamma_{1v}^{WZ} & \sim & z
\label{eq:WZ_bands_sym}
\end{eqnarray}

Although the IRs belong to different symmetry groups ($T_d$ and $C_{6v}$), the 
basis states transform as the usual $x,y,z$ cartesian coordinates for the valence 
bands and the scalar $x^{2}+y^{2}+z^{2}$ for the conduction band in both crystal 
structures. This information is crucial to represent WZ and ZB with the same 
Hamiltonian and is the essential insight of our formulation.

%================================================================================

\section{III. Theoretical model}

\subsection{A. k$\cdot$p Hamiltonian}

\begin{figure}[h]
\includegraphics{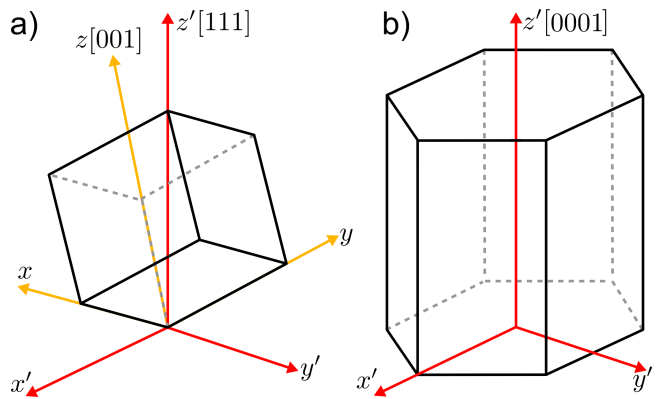}
\caption{(a) ZB conventional unit cell with two different coordinate systems.
(b) WZ conventional unit cell with its common coordinate system. The [111]
growth direction for ZB structure passes along the main diagonal of the cube
and is represented in the primed coordinate system.}
\label{fig:coordinate_systems}
\end{figure}

In order to develop our k$\cdot$p Hamiltonian \cite{arxiv} it is convenient to 
describe the ZB structure in a coordinate system that has the $z$ axis parallel 
to the growth direction. This coordinate system is the primed one presented in 
Figure \ref{fig:coordinate_systems}. Even though the choice of the coordinate 
system is arbitrary, it alters the k$\cdot$p Hamiltonian. For example, in the 
unprimed coordinate system the $k_z$ direction is directed towards the $X$-point 
but in the primed coordinate system it is directed towards the $L$-point. Thus 
our expectation is that an anisotropy in the ZB Hamiltonian between the 
$k_x$ and $k_z$ directions will occur since they are not reaching equivalent 
points in the the reciprocal space anymore.

Considering the single group formulation of the k$\cdot$p method, the choice of 
the coordinate system defines the symmetry operation matrices used to derive 
the momentum matrix elements. It is necessary to recalculate the Hamiltonian 
terms for the ZB structure. However, the energy bands we consider here are 
exactly the ones customary used in the ZB [001] k$\cdot$p Hamiltonian. Instead 
of recalculating the terms for ZB [111] k$\cdot$p Hamiltonian it is possible, 
and also useful, to apply a basis rotation to the ZB [001] matrix. This rotation 
procedure is well described in the paper of Park and Chuang \cite{jap-87-353}.

The basis set for both crystal structures in the primed coordinate system (the 
prime will be dropped out of the notation from now on and will be used only 
when it is necessary) is given by

\begin{eqnarray}
\left|c_{1}\right\rangle  & = & -\frac{1}{\sqrt{2}}\left|(X+iY)\uparrow\right\rangle \nonumber \\
\left|c_{2}\right\rangle  & = & \frac{1}{\sqrt{2}}\left|(X-iY)\uparrow\right\rangle \nonumber \\
\left|c_{3}\right\rangle  & = & \left|Z\uparrow\right\rangle \nonumber \\
\left|c_{4}\right\rangle  & = & \frac{1}{\sqrt{2}}\left|(X-iY)\downarrow\right\rangle \nonumber \\
\left|c_{5}\right\rangle  & = & -\frac{1}{\sqrt{2}}\left|(X+iY)\downarrow\right\rangle \nonumber \\
\left|c_{6}\right\rangle  & = & \left|Z\downarrow\right\rangle \nonumber \\
\left|c_{7}\right\rangle  & = & i\left|S\uparrow\right\rangle \nonumber \\
\left|c_{8}\right\rangle  & = & i\left|S\downarrow\right\rangle 
\end{eqnarray}

In a first approximation, the interband interaction is not taken into account explicitly 
here, thus the conduction band is a single band model for spin-up and spin-down 
reading as

\begin{equation}
E_{C}(\vec{k}) = E_{g}+E_{0}+\frac{\hbar^{2}}{2m_{e}^{\parallel}}k_{z}^{2}+\frac{\hbar^{2}}{2m_{e}^{\perp}}\left(k_{x}^{2}+k_{y}^{2}\right)
\end{equation}
\\
where $E_{g}$ is the band gap, $E_{0}$ is the energy reference at $\vec{k}=0$ and 
$m_{e}^{\parallel}$, $m_{e}^{\perp}$ are the electron effective masses parallel 
and perpendicular do the $z$ axis, respectively. For the ZB structure, however, 
the electron effective masses are equal.

The Hamiltonian for WZ and ZB valence band is given by

\begin{equation}
H_{V}(\vec{k}) = \left[\begin{array}{cccccc}
F & -K^{*} & -H^{*} & 0 & 0 & 0\\
-K & G & H & 0 & 0 & \Delta\\
-H & H^{*} & \lambda & 0 & \Delta & 0\\
0 & 0 & 0 & F & -K & H\\
0 & 0 & \Delta & -K^{*} & G & -H^{*}\\
0 & \Delta & 0 & H^{*} & -H & \lambda
\end{array}\right]
\label{eq:Hv_kp}
\end{equation}
\\
and the matrix terms are defined as

\begin{eqnarray}
F & = & \Delta_{1}+\Delta_{2}+\lambda+\theta\nonumber \\
G & = & \Delta_{1}-\Delta_{2}+\lambda+\theta\nonumber \\
\lambda & = & A_{1}k_{z}^{2}+A_{2}\left(k_{x}^{2}+k_{y}^{2}\right)\nonumber \\
\theta & = & A_{3}k_{z}^{2}+A_{4}\left(k_{x}^{2}+k_{y}^{2}\right)\nonumber \\
K & = & A_{5}k_{+}^{2}+2\sqrt{2}A_{z}k_{-}k_{z}\nonumber \\
H & = & A_{6}k_{+}k_{z}+A_{z}k_{-}^{2}\nonumber \\
\Delta & = & \sqrt{2}\Delta_{3}
\end{eqnarray}
\\
where $k_{\alpha}(\alpha=x,y,z)$ are the wave vectors in the primed coordinate 
system, $A_{i}(i=1,...,6,z)$ are the holes effective mass parameters, $\Delta_1$ 
is the crystal field splitting energy in WZ, $\Delta_{2,3}$ are the spin-orbit 
coupling splitting energies and $k_{\pm} = k_x \pm i k_y$. 

It is important to notice that the parameter $A_z$ appears in the matrix elements 
to regain the original isotropic symmetry of the ZB band structure in the new 
coordinate system. In the regions of WZ crystal structure, this parameter is 
zero and the matrix is exactly the canonical in use for WZ crystals.

Although this is not the usual way to describe ZB crystals, all the parameters 
in the matrix can be related to the familiar $\gamma_{i}(i=1,2,3)$ and 
$\Delta_{SO}$ as shown above

\begin{eqnarray}
\Delta_{1} & = & 0\nonumber \\
\Delta_{2} & = & \Delta_{3}=\frac{\Delta_{SO}}{3}\nonumber \\
A_{1} & = & -\gamma_{1}-4\gamma_{3}\nonumber \\
A_{2} & = & -\gamma_{1}+2\gamma_{3}\nonumber \\
A_{3} & = & 6\gamma_{3}\nonumber \\
A_{4} & = & -3\gamma_{3}\nonumber \\
A_{5} & = & -\gamma_{2}-2\gamma_{3}\nonumber \\
A_{6} & = & -\sqrt{2}\left(2\gamma_{2}+\gamma_{3}\right)\nonumber \\
A_{z} & = & \gamma_{2}-\gamma_{3}
\end{eqnarray}

One may also argue that this formulation is very similar to the WZ phase. 
The insight here is to consider ZB as a WZ structure without the crystal 
field splitting energy. Since the WZ structure is less symmetric than ZB, 
as mentioned in section II.A, it is possible to represent the ZB 
parameters with the WZ ones.

\begin{figure}[h]
\begin{center}
\includegraphics{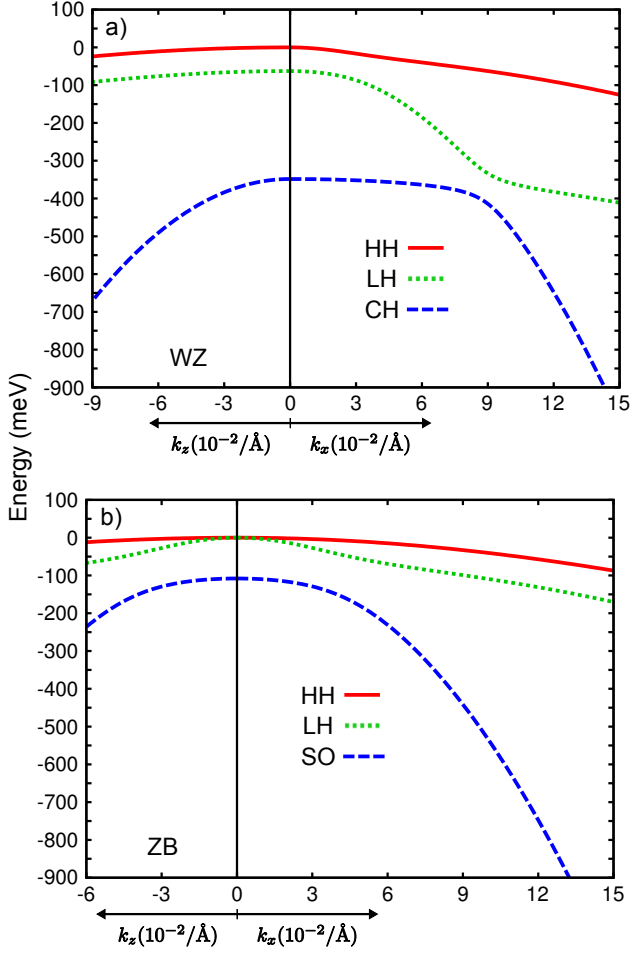}
\caption{Valence band structure for bulk (a) WZ and (b) ZB in the primed coordinate
system. The usual identification of the bands was used for ZB while in WZ it was
necessary to analyze the composition of the states in the $\Gamma$-point. We can
see the anisotropy between $k_z$ and $k_x$ in WZ and also in ZB because in the
new coordinate system the $x$ and $z$ axes do not reach equivalent points in
the reciprocal space. The top of the valence band in both crystal structures was
chosen to be at zero.}
\label{fig:valence_band_bulk}
\end{center}
\end{figure}

The resulting valence band structures for bulk WZ and ZB InP using matrix 
(\ref{eq:Hv_kp}) are shown in Figure \ref{fig:valence_band_bulk}. The presence of the 
crystal field in the WZ structure creates three distinct two-fold degenerate 
bands whereas in ZB there is a four-fold and a two-fold degenerate set of 
bands. Additionally, the anisotropy between $k_x$ and $k_x$ is evident in 
both crystal structures. In WZ it is due to the different symmetry properties 
of the $xy$-plane and the $z$ axis but in ZB it is because $k_x$ and $k_z$ 
directions do not reach equivalent points in the reciprocal space. Since the 
conduction band is a parabolic model, we did not present its dispersion 
relation. The ZB parameters were obtained from Ref. \cite{jap-89-5815} and the 
WZ parameters were derived using the effective masses presented in Ref. 
\cite{prb-81-155210}. These parameters can be found in Table \ref{tab:kp_par}.

% HH - solid line
% LH - dotted line
% CH ou SO - dashed line

Following Chuang and Chang \cite{apl-68-1657} notation, the valence energy bands 
for WZ are named after the composition of the states at $\vec{k}=0$. HH (heavy 
hole) states are composed only by $\left|c_{1}\right\rangle$ or 
$\left|c_{4}\right\rangle$, LH (light hole) states are composed mainly of 
$\left|c_{2}\right\rangle$ or $\left|c_{5}\right\rangle$ and CH (crystal-field 
split-off hole) are composed mainly of $\left|c_{3}\right\rangle$ or 
$\left|c_{6}\right\rangle$. For the ZB structure the common identification of 
the valence energy bands was used. The four-fold degenerate bands at $\vec{k}=0$ 
are HH and LH and the lower two-fold degenerate band is SO (split-off hole).

\begin{table}[h]
\begin{center}
\caption{InP parameters used in the calculations.}
%\resizebox{8.5cm}{!} 
%{
\begin{tabular}{ccc}
\hline 
\hline 
Parameter & ZB InP & WZ InP\tabularnewline
\hline 
Lattice constant ($\textrm{\AA}$) &  & \tabularnewline
$a$ & 5.8697 & 4.1505 \tabularnewline
$c$ & - & 6.7777 \tabularnewline
Energy parameters (eV) &  & \tabularnewline
$E_{g}$ & 1.4236 & 1.474\tabularnewline
$\Delta_{1}$ & 0 & 0.303\tabularnewline
$\Delta_{2}=\Delta_{3}$ & 0.036 & 0.036\tabularnewline
Conduction band effective masses &  & \tabularnewline
$m_{e}^{\parallel}/m_{0}$ & 0.0795 & 0.105\tabularnewline
$m_{e}^{\perp}/m_{0}$ & 0.0795 & 0.088\tabularnewline
Valence band effective mass \\ parameters (units of $\frac{\hbar^{2}}{2m_{0}}$) &  & \tabularnewline
$A_{1}$ & -13.4800 & -10.7156 \tabularnewline
$A_{2}$ & -0.8800 & -0.8299 \tabularnewline
$A_{3}$ & 12.6000 & 9.9301 \tabularnewline
$A_{4}$ & -6.3000 & -5.2933 \tabularnewline
$A_{5}$ & -5.8000 & 5.0000 \tabularnewline
$A_{6}$ & -7.4953 & 1.5000 \tabularnewline
$A_{z}$ & -0.5000 & 0 \tabularnewline
\hline 
\hline 
\label{tab:kp_par}
\end{tabular}
%}
\end{center}
\end{table}

%--------------------------------------------------------------------------------

\subsection{B. Strain}

The strain Hamiltonian can be obtained using the same basis rotation applied to 
the k$\cdot$p matrix \cite{jap-87-353}. Similarly, the conduction and valence 
band are decoupled.

For the conduction band, the strain effect is given by

\begin{equation}
E_{C\varepsilon} = a_{c\parallel}\varepsilon_{zz}+a_{c\perp}\left(\varepsilon_{xx}+\varepsilon_{yy}\right)
\end{equation}
\\
where $a_{c\parallel}$ and $a_{c\perp}$ are the conduction band deformation potentials parallel and perpendicular to 
the $z$ axis, respectively. In the ZB structure they have the same value.

The valence band strain Hamiltonian is

\begin{equation}
H_{V\varepsilon} = \left[\begin{array}{cccccc}
F_{\varepsilon} & -K_{\varepsilon}^{*} & -H_{\varepsilon}^{*} & 0 & 0 & 0\\
-K_{\varepsilon} & F_{\varepsilon} & H_{\varepsilon} & 0 & 0 & 0\\
-H_{\varepsilon} & H_{\varepsilon}^{*} & \lambda_{\varepsilon} & 0 & 0 & 0\\
0 & 0 & 0 & F_{\varepsilon} & -K_{\varepsilon} & H_{\varepsilon}\\
0 & 0 & 0 & -K_{\varepsilon}^{*} & F_{\varepsilon} & -H_{\varepsilon}^{*}\\
0 & 0 & 0 & H_{\varepsilon}^{*} & -H_{\varepsilon} & \lambda_{\varepsilon}
\end{array}\right]
\label{eq:Hv_strain}
\end{equation}
\\
and the matrix terms are 

\begin{eqnarray}
F_{\varepsilon} & = & \left(D_{1}+D_{3}\right)\varepsilon_{zz}+\left(D_{2}+D_{4}\right)\left(\varepsilon_{xx}+\varepsilon_{yy}\right)\nonumber \\
\lambda_{\varepsilon} & = & D_{1}\varepsilon_{zz}+D_{2}\left(\varepsilon_{xx}+\varepsilon_{yy}\right)\nonumber \\
K_{\varepsilon} & = & D_{5}^{(1)}\left(\varepsilon_{xx}-\varepsilon_{yy}\right)+D_{5}^{(2)}2i\varepsilon_{xy}\nonumber \\
H_{\varepsilon} & = & D_{6}\left(\varepsilon_{xz}+i\varepsilon_{yz}\right)+D_{z}\left(\varepsilon_{xx}-\varepsilon_{yy}\right)
\end{eqnarray}
\\
where $D_i$'s are the valence band deformation potentials and $\varepsilon_{ij}$ 
is the strain tensor.

In the same way as the $A_{z}$ parameter appears in k$\cdot$p matrix, some extra 
deformation potential terms were appears to use the same strain Hamiltonian 
for both crystal structures. The deformation potential $D_{5}$ was split in two 
parts because the strain tensor $\varepsilon_{xy}$ is not present in the ZB 
structure. For WZ, $D^{(1)}_{5}=D^{(2)}_{5}$. Also, the $D_{z}$ deformation 
potential takes into account the non existing term 
$\varepsilon_{xx}-\varepsilon_{yy}$ in the WZ structure.

The deformation potentials $D_i$'s are related to the ZB ones

\begin{eqnarray}
D_{1} & = & a_{v}+\frac{2d}{\sqrt{3}}\nonumber \\
D_{2} & = & a_{v}-\frac{d}{\sqrt{3}}\nonumber \\
D_{3} & = & -\sqrt{3}d\nonumber \\
D_{4} & = & \frac{3d}{2\sqrt{3}}\nonumber \\
D_{5}^{(1)} & = & -\frac{b}{2}-\frac{d}{\sqrt{3}}\nonumber \\
D_{5}^{(2)} & = & 0\nonumber \\
D_{6} & = & 0\nonumber \\
D_{z} & = & -\frac{b}{2}+\frac{d}{2\sqrt{3}}\nonumber \\
a_{c\parallel} & = & a_{c\perp}=a_{c}
\end{eqnarray}

Considering biaxial strain, the elements of the strain tensor can be obtained in 
both coordinate systems for ZB and WZ. Although it is convenient to describe the 
Hamiltonian terms in the primed coordinate system, it is also useful to describe 
the strain tensor elements in the unprimed coordinate system. They will be used 
to construct the piezoelectric polarization in section III.C. The prime will be 
reintegrated in the notation to avoid confusion and the upper scripts $z$ and $w$ 
denotes the ZB and WZ structures, respectively.

For the primed coordinate system, the elements of the strain tensor are given by

\begin{equation}
\varepsilon_{xx}^{\prime(z,w)} = \varepsilon_{yy}^{\prime(z,w)}=\frac{a_{0}-a^{(z,w)}}{a^{(z,w)}}
\end{equation}

\begin{equation}
\varepsilon_{zz}^{\prime(z)} = -\frac{1}{\sigma^{(111)}} \varepsilon_{xx}^{\prime(z)}
\label{eq:prime_z}
\end{equation}

\begin{equation}
\varepsilon_{zz}^{\prime(w)} = -\frac{2C^{(w)}_{13}}{C^{(w)}_{33}}\varepsilon_{xx}^{\prime(w)}
\label{eq:prime_w}
\end{equation}

\begin{equation}
\varepsilon_{yz}^{\prime(z,w)} = \varepsilon_{zx}^{\prime(z,w)}=\varepsilon_{xy}^{\prime(z,w)} = 0
\end{equation}
\\
where $a_{0}$ is the lattice constant of the substrate.

In the unprimed coordinate system, the strain tensor elements assume the form

\begin{equation}
\varepsilon_{xx}^{(z)} = \varepsilon_{yy}^{(z)} = \varepsilon_{zz}^{(z)} = \frac{1}{3}\left(2-\frac{1}{\sigma^{(111)}}\right)\varepsilon_{xx}^{\prime(z)}
\end{equation}

\begin{equation}
\varepsilon_{yz}^{(z)} = \varepsilon_{zx}^{(z)} = \varepsilon_{xy}^{(z)} = -\frac{1}{3}\left(1+\frac{1}{\sigma^{(111)}}\right)\varepsilon_{xx}^{\prime(z)}
\end{equation}

The quantity $\sigma^{(111)}$ is given by

\begin{equation}
\sigma^{(111)} = \frac{C^{(z)}_{11} + 2C^{(z)}_{12} + 4C^{(z)}_{44}}{2C^{(z)}_{11} + 4C^{(z)}_{12} - 4C^{(z)}_{44}}
\end{equation}

Comparing the expression (\ref{eq:prime_z}) with (\ref{eq:prime_w}) it is possible 
to obtain effective values for $C^{(z)}_{13}$ and $C^{(z)}_{33}$ for ZB in the 
primed coordinate system. The effective values are:

\begin{equation}
C^{(z)}_{13}=C^{(z)}_{11}+2C^{(z)}_{12}-2C^{(z)}_{44}
\end{equation}

\begin{equation}
C^{(z)}_{33}=C^{(z)}_{11}+2C^{(z)}_{12}+4C^{(z)}_{44}
\end{equation}

Thus, we have a single set of expressions to describe biaxial strain in the 
primed coordinate system for ZB and WZ crystal structures:

\begin{equation}
\varepsilon_{xx}^{\prime} = \varepsilon_{yy}^{\prime}=\frac{a_{0}-a}{a}
\end{equation}

\begin{equation}
\varepsilon_{zz}^{\prime} = -\frac{2C^{(z,w)}_{13}}{C^{(z,w)}_{33}}\varepsilon_{xx}^{\prime}
\end{equation}

\begin{equation}
\varepsilon_{yz}^{\prime} = \varepsilon_{zx}^{\prime} = \varepsilon_{xy}^{\prime} = 0
\end{equation}

Since deformation potentials and elastic stiffness constants for WZ InP are not 
yet available in the literature, we will consider here that the strain effect 
appears only in the ZB structure. This assumption is not totally unrealistic 
because WZ is the dominant phase in the NW.

\begin{figure}[h]
\begin{center}
\includegraphics{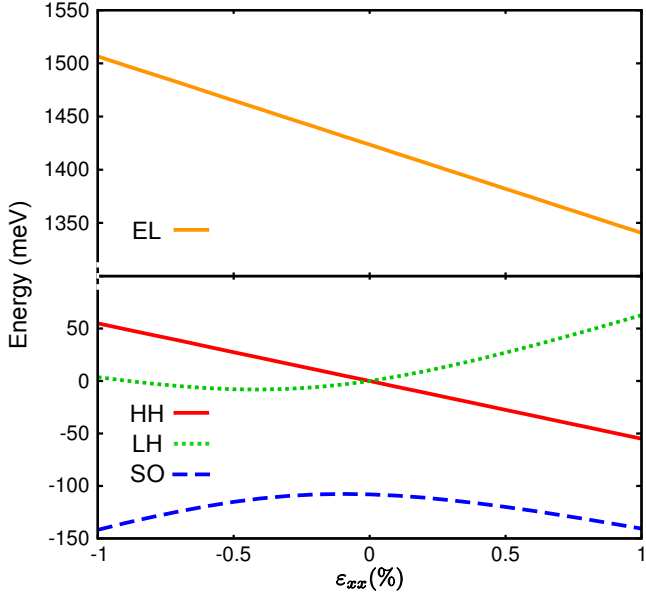}
\caption{Strain effect in the band edges of ZB InP as a function of the percentage
of strain tensor. EL and HH shows a linear variation while LH and SO shows a
nonlinear behavior. The strain effect removes the HH and LH degeneracy.}
\label{fig:strain_zb_edges}
\end{center}
\end{figure}
% EL - solid line
% HH - solid line
% LH - dotted line
% SO - dashed line

Figure \ref{fig:strain_zb_edges} shows the effect of strain at $\vec{k}=0$ for 
the diagonalized Hamiltonian (k$\cdot$p and strain terms) as a function of the 
percentage of strain tensor. A linear variation for the conduction band and the 
HH band is observed, however, the LH and SO bands have a non-linear behavior. The 
order of the HH and LH bands changes when strain is distensive. Table 
\ref{tab:strain_par} lists the ZB parameters used in the calculations.

\begin{table}[h]
\begin{center}
\caption{ZB InP strain parameters.}
%\resizebox{8.5cm}{!} 
%{
\begin{tabular}{cc}
\hline 
\hline 
Parameter & ZB InP\tabularnewline
\hline 
Deformation potentials (eV) & \tabularnewline
$D_{1}$ & -6.3735 \tabularnewline
$D_{2}$ & 2.2868 \tabularnewline
$D_{3}$ & 8.6603 \tabularnewline
$D_{4}$ & -4.3301 \tabularnewline
$D^{(1)}_{5}$ & 3.8868 \tabularnewline
$D^{(2)}_{5}$ & 0 \tabularnewline
$D_{6}$ & 0 \tabularnewline
$D_{z}$ & -0.4434 \tabularnewline
$a_{c\par}$ & -6.0 \tabularnewline
$a_{c\perp}$ & -6.0 \tabularnewline
Elastic stiffness constant (GPa) & \tabularnewline
$C_{11}$ & 1011 \tabularnewline
$C_{12}$ & 561 \tabularnewline
$C_{44}$ & 456 \tabularnewline
\hline 
\hline 
\label{tab:strain_par}
\end{tabular}
%}
\end{center}
\end{table}

%--------------------------------------------------------------------------------

\subsection{C. Spontaneous and Piezoelectric Polarization}

Piezoelectric polarization appears when a crystal is subjected to strain. In ZB 
semiconductors grown along the [111] direction, the magnitude of the piezoelectric 
polarization, in the unprimed coordinate system of Fig. \ref{fig:coordinate_systems}, 
is given by \cite{prb-35-1242}:

\begin{equation}
P_{i} = 2e_{14}\varepsilon_{jk}
\end{equation}
\\
where $e_{14}$ is the piezoelectric constant for ZB materials, $(i,j,k)$ are the 
cartesian coordinates $(x,y,z)$ in a cyclic order and $\varepsilon_{jk}$ are the 
strain tensor components.

Applying the coordinate system rotation in the piezoelectric polarization vector 
components in order to describe them in the primed coordinate system we obtain:

\begin{eqnarray}
P_{x}^{\prime} & = & \frac{1}{\sqrt{6}}\left(P_{x}+P_{y}-2P_{z}\right)=0\nonumber \\
P_{y}^{\prime} & = & \frac{1}{\sqrt{2}}\left(-P_{x}+P_{y}\right)=0\nonumber \\
P_{z}^{\prime} & = & \frac{1}{\sqrt{3}}\left(P_{x}+P_{y}+P_{z}\right)=\sqrt{3}P
\end{eqnarray}

The resulting piezoelectric polarization alongside the growth direction is then:

\begin{equation}
P_{z}^{\prime} = -\frac{2}{\sqrt{3}}e_{14}\left(1+\frac{1}{\sigma^{(111)}}\right)\varepsilon_{xx}^{\prime}
\end{equation}

The spontaneous polarization effect in WZ structure is due to the relative 
displacement between the cations and anions when the ratio $c/a$ is different 
from the ideal value in the WZ structure.

In a heterostructure, the effect of the different polarizations in each region 
creates an electric field through the whole structure. The net electric field in 
a determined layer, $i$, due to spontaneous and piezoelectric polarizations in 
the system is given by \cite{paul-harrison}:

\begin{equation}
E_{i}=\frac{\overset{N}{\underset{j=1}{\sum}}\left(P_{j}-P_{i}\right)\frac{l_j}{\varepsilon_j}}{\varepsilon_{i}\overset{N}{\underset{j=1}{\sum}}\frac{l_j}{\varepsilon_j}}
\end{equation}
\\
where $j$ sums all over the layers in the heterostructure with polarization $P$, 
dielectric constant $\varepsilon$ and length $l$.

%--------------------------------------------------------------------------------

\subsection{D. Effective mass equation in reciprocal space}

The envelope function approximation \cite{book-bastard, IEEEjqe-22-1625} is applied 
to couple the different crystal structures alongside the growth direction in the NW. 
In each region the wave function is expanded in terms of the Bloch functions of the 
corresponding polytype. Thus, the wave function of the whole system is given by:

\begin{equation}
\psi(\vec{r}) = \sum_{l}e^{i(\vec{k}\cdot\vec{r})}g_{l}(\vec{r})u_{l}^{(WZ,ZB)}(\vec{r})
\label{eq:eva}
\end{equation}
\\
where $g_{l}(\vec{r})$ are the envelope functions of the $l$-th basis state.

Considering different Bloch functions for each region, the Hamiltonian parameters 
vary alongside the growth direction, making it possible to use the common k$\cdot$p 
and strain matrices, (\ref{eq:Hv_kp}) and (\ref{eq:Hv_strain}), for both crystal 
structures. Moreover, since each crystal structure dictates its symmetry to their 
respective Bloch functions, some matrix elements can be forbidden by symmetry in the 
region of a certain crystalline phase. For example, the $A_z$ parameter is zero in 
WZ regions whereas the $\Delta_1$ parameter is zero in ZB regions.

To represent the growth dependence of the Hamiltonian parameters and envelope functions, 
the plane wave expansion is used. This formalism considers the periodicity of the whole 
system allowing the expansion of growth dependent functions in Fourier coefficients:

\begin{equation}
U(\vec{r}) = \sum_{\vec{K}} U_{\vec{K}} e^{i \vec{K} \cdot \vec{r}}
\label{eq:pwe}
\end{equation}
\\
where $U_{\vec{K}}$ are the Fourier coefficients of the function $U(\vec{r})$ and 
$\vec{K}$ is a reciprocal lattice vector. The Fourier expansion also induces the 
change $\vec{k} \rightarrow \vec{k} + \vec{K}$ in the k$\cdot$p matrix.

%--------------------------------------------------------------------------------

%================================================================================

\section{IV. Results and discussion}

The NW system chosen to apply our model is a WZ/ZB/WZ single well structure. Although 
a real NW is composed of multiple polytypical quantum wells with different sizes, the 
analysis of just a single well can bring out the physics of the polytypical interface. 
The effects of lateral confinement are neglected in a first approach, assuming NWs with 
large lateral dimensions. Strain, piezoelectric and spontaneous polarization are also 
included in the single well system.

\begin{figure}[h]
\begin{center}
\includegraphics{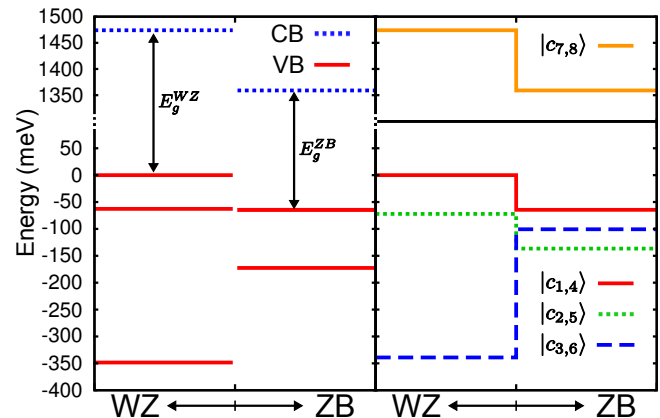}
\caption{Left side: Band edge energies at $\vec{k}=0$ for the diagonalized Hamiltonian.
Right side: Diagonal terms of the Hamiltonian at $\vec{k}=0$.}
\label{fig:kp_interface}
\end{center}
\end{figure}

When both crystal structures are put side by side, a band-offset is created at the 
interface, originating a confinement profile. The band mismatch is also taken from 
reference \cite{prb-81-155210}. Figure \ref{fig:kp_interface} exhibits the WZ/ZB 
interface for InP in two different schemes: in the left, the energies in $\vec{k}=0$ 
for the diagonalized Hamiltonian and in the right, the diagonal terms of the Hamiltonian. 
Although the composition of the states in the diagonalized energy bands are the same 
in $\vec{k}=0$, the matrix is not constructed in this basis. Since the variation along 
the growth direction of the matrix elements is well defined, the scheme in the right 
is more convenient to analyze the potential profile of the system.

\begin{figure}[h]
\begin{center}
\includegraphics{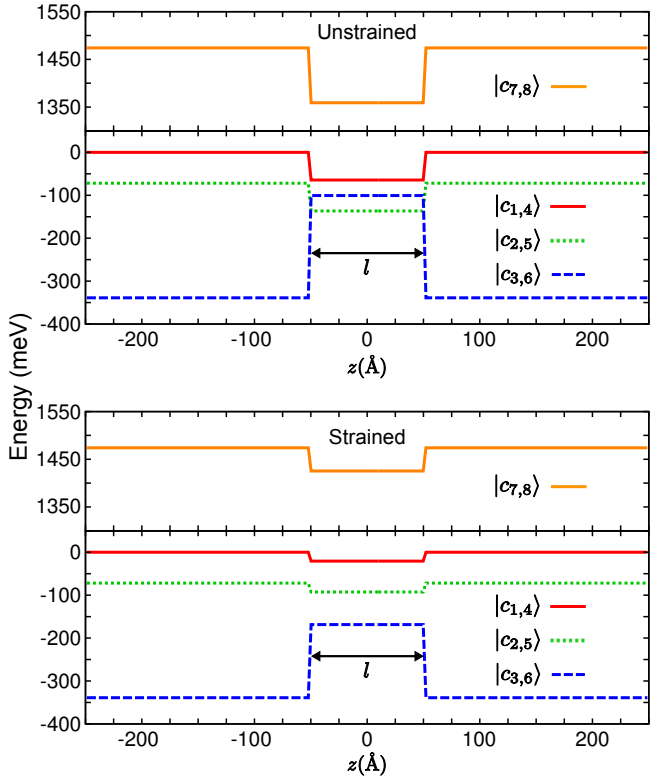}
\caption{Diagonal potential profile of the Hamiltonian for the polytypical InP system
with and without strain. The width of the ZB region, $l$, can change but the whole
system's dimension remains constant with $500\,\textrm{\AA}$.}
\label{fig:single_well_and_strain_pots}
\end{center}
\end{figure}

In all performed calculations, the entire length of the system is set to 
$500\,\textrm{\AA}$ and the width of the ZB region, $l$, is variable. Figure 
\ref{fig:single_well_and_strain_pots} shows the single well potentials with and 
without the effect of strain. Since the potential profiles exhibits a type-II behavior, 
we expect a spatial separation of the carriers: electrons are more likely to be in the 
ZB region and the holes in the WZ region.

The strain considered here, $-0.8\%$, is a intermediate value between two data 
available in the literature from \textit{ab initio} calculations. Reference 
\cite{ssc-151-781} shows that the deviation between the lattice constant of ZB[111] 
and WZ[0001] is $-1.3\%$ and reference \cite{nanotec-21-505709} shows $-0.3\%$. Also, 
reference \cite{nanolet-10-4055} suggests a difference slighter than $0.5\%$ between 
the lattice constants of the two polytypes. The effect of strain shallows the potential 
wells in the conduction and valence bands, reducing the confinement of the carriers. We 
expect to have less confined states for the strained potential compared to the unstrained one. 

\begin{figure}[h]
\begin{center}
\includegraphics{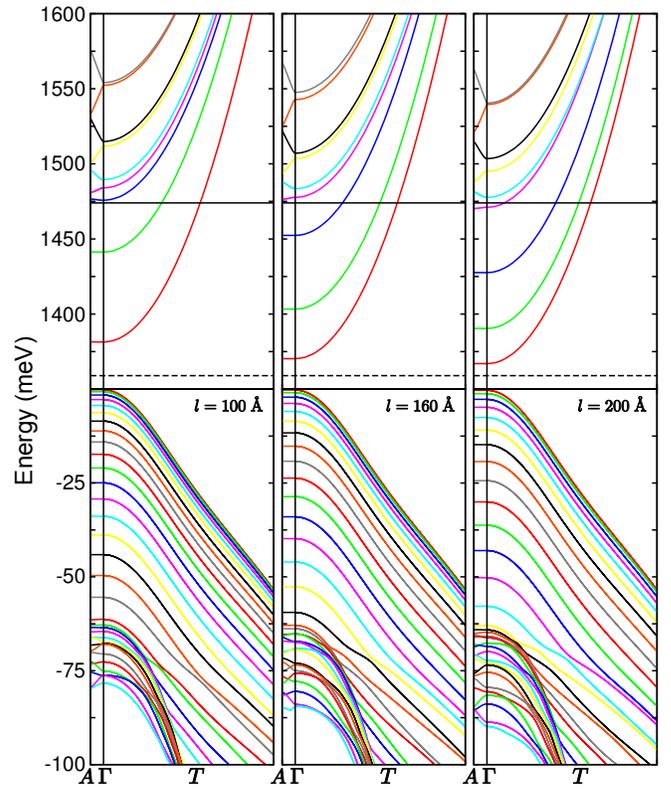}
\caption{Band structure for the unstrained profile in Figure \ref{fig:single_well_and_strain_pots}.
The solid horizontal line is the top energy of conduction band well and the dashed horizontal line
is the bottom energy of the conduction band well. The $\Gamma-T$ direction refers to $k_x$ and 
$\Gamma-A$ to $k_z$.}
\label{fig:single_well_bs}
\end{center}
\end{figure}

The conduction and valence band structures for the potential profile without strain 
are presented in Fig. \ref{fig:single_well_bs} for three different widths of the ZB region. 
The calculations were performed up to 10\% in the $\Gamma-T$ direction and 100\% in 
the $\Gamma-A$ direction. For the valence band 64 energy states are presented while 
only 18 are presented for the conduction band. Since the system has no asymmetric 
potential, the energy bands are two-fold degenerate in spin, therefore 32 states are 
visible in the valence band and 9 in the conduction band. For the three different 
values of $l$, the conduction energy bands are nominated, from bottom to top, as 
EL1-EL9, composed of $\left|c_{7,8}\right\rangle$ states. The valence bands, from 
top to bottom, are nominated as HH1-19, LH1-4, HH20-21, LH5-7, HH22-23, LH8-9 for 
$l = 100\,\textrm{\AA}$; HH1-16, LH1-2, HH17, LH3, HH18, LH4, HH19, LH5-7, HH20-21, 
LH8-9, HH22-23 for $l = 160\,\textrm{\AA}$ and HH1-14, LH1-2, HH15, LH3, HH16-17, 
LH4-5, HH18, LH6, HH19, LH7, HH20, LH8, HH21, LH9, HH22, LH10 for $l = 200\,\textrm{\AA}$.

Since the highest valence band states are HH there is no significant anticrossing among the 
energy bands in the $\Gamma-T$ direction. The anticrossing is characteristic of interactions 
between HH and LH bands in ZB and WZ quantum well structures. A slight anticrossing, however, 
can be seen in the energy region just above $-75\,\text{meV}$, which is next to the interaction 
region of the $\left|c_{1,4}\right\rangle$ and $\left|c_{2,5}\right\rangle$ profiles.

Increasing the value of $l$ we find that the number of confined states in the conduction band 
increases. On the other hand, for the valence band the number of confined states decreases 
because the WZ region's width also decreases.

\begin{figure}[h]
\begin{center}
\includegraphics{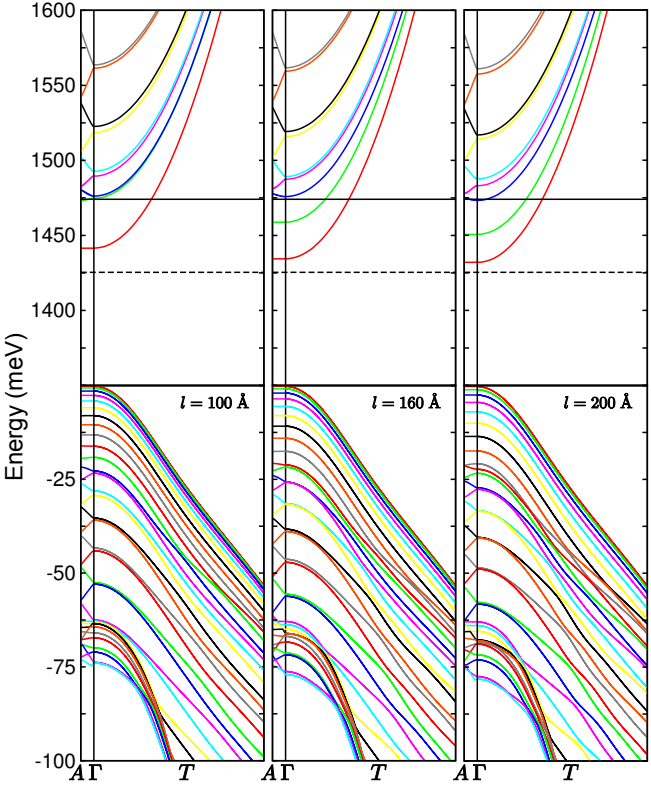}
\caption{Band structure for the strained profile in Fig. \ref{fig:single_well_and_strain_pots}.
The solid and dashed lines have the same meaning as in Figure \ref{fig:single_well_bs}. The
$\Gamma-T$ direction refers to $k_x$ and $\Gamma-A$ to $k_z$.}
\label{fig:single_well_strain_bs}
\end{center}
\end{figure}

For the strained potential profile, the band structures for three different widths of 
the ZB region are displayed in Figure \ref{fig:single_well_strain_bs}. The calculations were 
performed considering the same extension for the FBZ of the unstrained band structure. 
For the three different values of $l$, the conduction energy bands are nominated, from 
bottom to top, as EL1-EL9, composed of $\left|c_{7,8}\right\rangle$ states. The valence 
bands, from top to bottom, are nominated as HH1-22, LH1, HH23, LH2-7, HH24-25 for 
$l = 100\,\textrm{\AA}$; HH1-21, LH1-3, HH22-23, LH4-8, HH24  for $l = 160\,\textrm{\AA}$ 
and HH1-21, LH1-4, HH22-23, LH5-9 for $l = 200\,\textrm{\AA}$.

As expected, the number of confined states for the strained profile compared to the 
unstrained one is smaller. Nonetheless, the similar confinement trend is visible here 
when the value of $l$ increases: the number confined states in the conduction band 
increases while in valence band decreases.

An interesting feature presented in the strained band structure is the presence of some 
confined states below the top region, around $-62\,\text{meV}$. This suggests a confinement 
in the intermediate region of the $\left|c_{2,5}\right\rangle$ and $\left|c_{3,6}\right\rangle$ 
profiles. Note that the coupling of these two profiles at $\vec{k}=0$ happens because of 
the off-diagonal spin-orbit term.

The composition of the energy states at $\vec{k}=0$ in the band structure is similar for 
the strained and unstrained cases: they are just HH or LH states. There is no CH states 
in the energy range considered here. The major contribution for CH states comes from 
the $\left|c_{3,6}\right\rangle$ profile, which is the lowest one in both cases.

The information of the energy states' composition can reveal important trends in the 
luminescence spectra for this kind of system. For example, at $\vec{k}=0$ the dominant 
symmetry of the energy states belongs to $\left(x,y\right)$, which means that the 
luminescence spectra is more intense perpendicular to the growth direction. However, 
experimental measurements \cite{prb-82-125327} indicates that the intensities 
perpendicular and parallel to the growth direction are almost similar. Therefore, we expect 
that the contribution for the parallel luminescence comes from the states at $\vec{k} \neq 0$. 
For $\vec{k}$ points away from the $\Gamma$-point, there is a stronger mixing between all 
the basis states.

\begin{figure}[h]
\begin{center}
\includegraphics{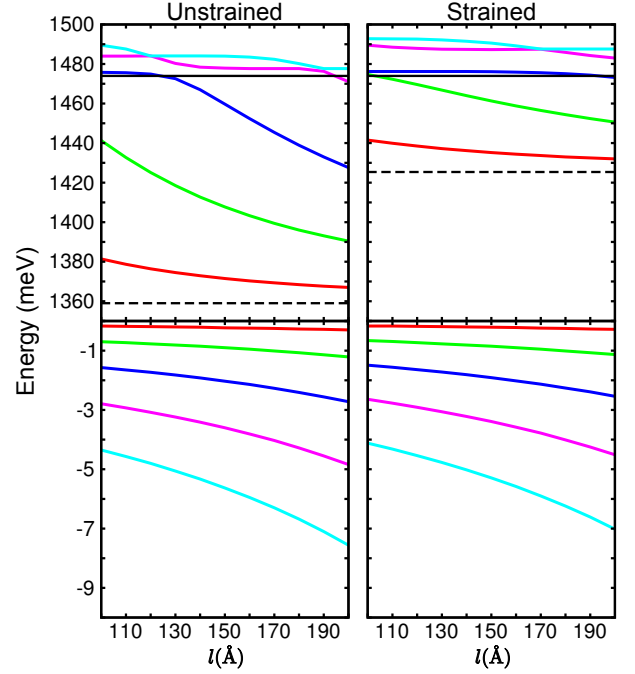}
\caption{The first 5 states of the conduction and valence bands at $\vec{k}=0$ as
a function of the ZB region width $l$. The solid line indicates the top of the conduction
band well and the dashed line indicates the bottom.}
\label{fig:single_well_and_strain_gamma_states}
\end{center}
\end{figure}

The effect of the ZB region width for both strained and unstrained potential profile 
for the conduction and valence band states at $\vec{k}=0$ is presented in Figure 
\ref{fig:single_well_and_strain_gamma_states}. It is possible to observe that the number of 
confined states in the conduction band increases as the value of $l$ increases. On the other hand, 
the number of confined states in the valence band decreases. Nevertheless, the effect of the 
variation of $l$ is more significant for the conduction band since the electron effective mass in 
ZB ($m_e^*/m_0=0.0795$) is smaller than the heavy hole mass of WZ ($m_{HH}^\parallel/m_0=1.273$ and 
$m_{HH}^\perp/m_0=0.158$). The same trend is also observed in the strained case. Also, since the 
bottom of the well in the conduction band has a higher value in the strained case, we can expect the 
interband transition energies to be blueshifted with the inclusion of strain effects.

The presence of strain effects gives rise to the piezoelectric polarization. For the InP ZB, 
the value for the piezoelectric constant used was $e_{14}=0.035\,\text{C/m}^2$, taken from Ref. 
\cite{jap-92-932}. For both crystal structures, the value used for static dielectric constant 
was $12.5$. The unknown parameter is the spontaneous polarization for WZ InP. Ref. \cite{nanolet-10-4055} 
suggests that this value is smaller than that of InN ($-0.03\,C/m^2$). In an attempt to 
estimate this value for InP, we performed the band structure calculations considering 
a range of values for $P_{sp}$.

\begin{figure}[h]
\begin{center}
\includegraphics{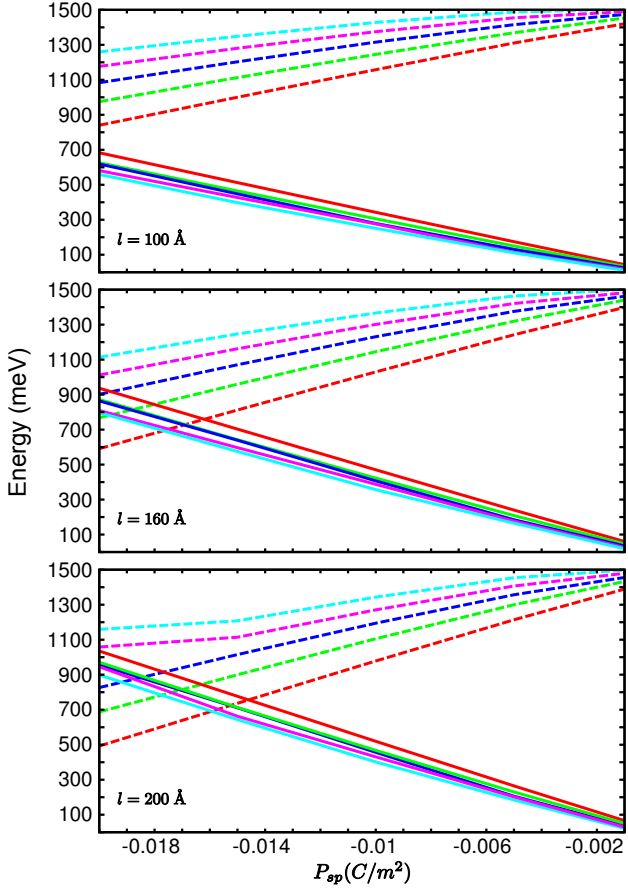}
\caption{The first 5 states of the conduction and valence bands at $\vec{k}=0$ as
a function of WZ spontaneous polarization $P_{sp}$. Notice the crossing of valence 
and conduction bands.}
\label{fig:sp_gamma_states}
\end{center}
\end{figure}

The energy of the first 5 conduction and valence band states at $\vec{k}=0$ as a function 
of spontaneous polarization in WZ InP for three different ZB region widths is presented 
in Figure \ref{fig:sp_gamma_states}. The considered values for spontaneous polarization 
are $-0.02\,C/m^2$, $-0.015\,C/m^2$, $-0.01\,C/m^2$, $-0.005\,C/m^2$ and $-0.001\,C/m^2$. For 
$l=160\,\textrm{\AA}$ and $l=200\,\textrm{\AA}$ there is a crossing between the conduction 
and valence band states. This is not observed experimentally therefore we consider this 
region \textit{forbidden}. Then, the \textit{allowed} values for spontaneous polarization 
considered here are then $-0.01\,C/m^2$, $-0.005\,C/m^2$ and $-0.001\,C/m^2$.

\begin{figure}[h]
\begin{center}
\includegraphics{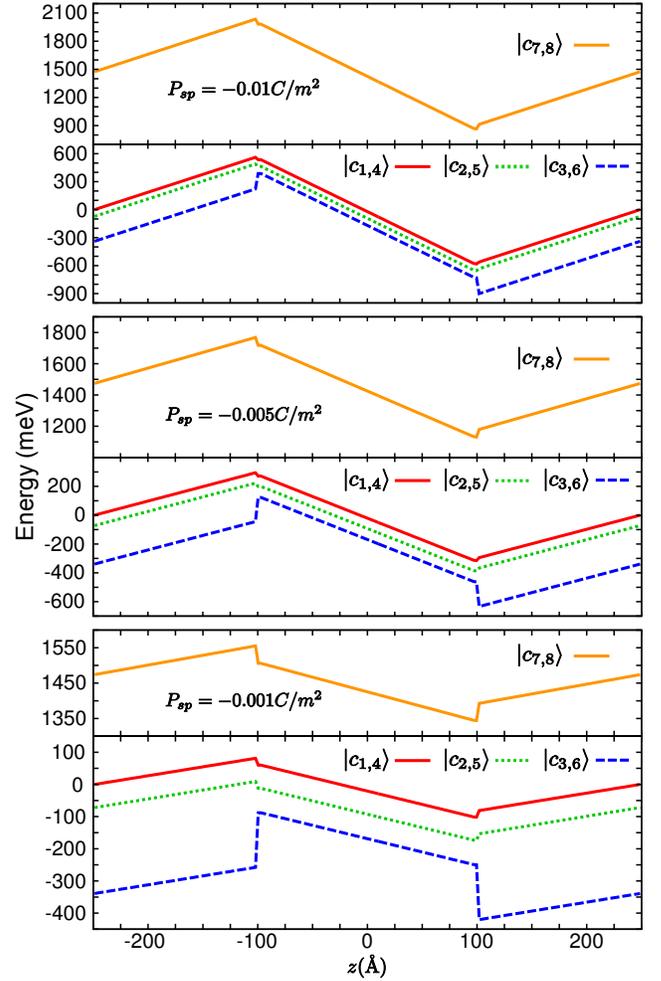}
\caption{Diagonal potential profile of the Hamiltonian for the polytypical InP system
considering strain, piezoelectric polarization in the ZB region and spontaneous
polarization in the WZ region for $l=200\,\textrm{\AA}$. The values for the spontaneous
polarization were chosen from Figure \ref{fig:sp_gamma_states}.} 
\label{fig:sppz_well_200_pots}
\end{center}
\end{figure}

The diagonal potential profile including effects of piezoelectric and spontaneous polarization 
is presented in Figure \ref{fig:sppz_well_200_pots}. The ZB region is fixed at $200\,\textrm{\AA}$. 
Analyzing these profiles, we expect to have strong coupling in the band structure for higher values 
of $P_{sp}$ since the profiles are more close to each other. This induces the mixing of states because 
an energy value can be in more than one profile.

\begin{figure}[h]
\begin{center}
\includegraphics{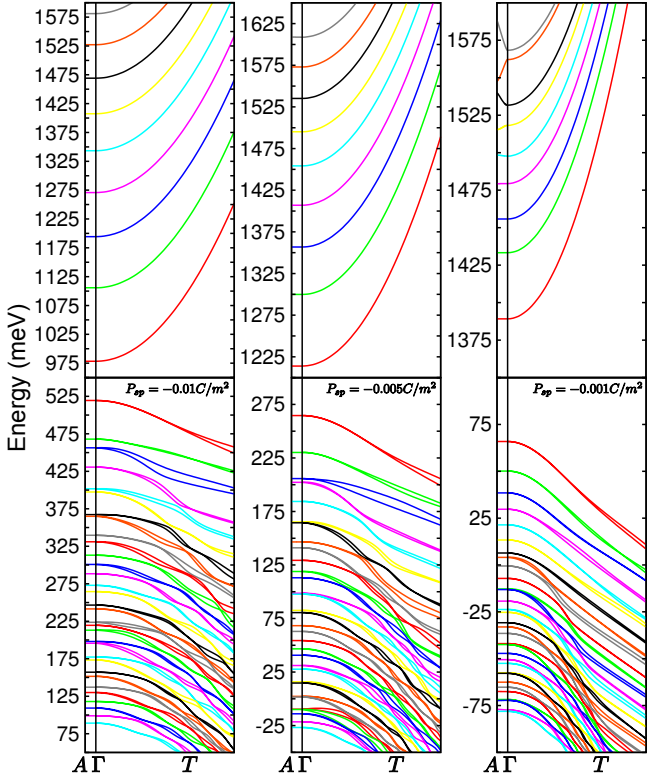}
\caption{Band structure for the profiles presented in Figure \ref{fig:sppz_well_200_pots}.
The spin-splitting of the energy bands is due to the field induced asymmetry.}
\label{fig:sppz_well_200_bs}
\end{center}
\end{figure}

The resulting band structures for the three different potential profiles of Figure 
\ref{fig:sppz_well_200_pots} are shown in Figure \ref{fig:sppz_well_200_bs}.
For the three different values of $P_{sp}$, the conduction energy bands are nominated, 
from bottom to top, as EL1-EL9, composed of $\left|c_{7,8}\right\rangle$ states. 
The valence bands, from top to bottom, are nominated as HH1-2, LH1, HH3, LH2, HH4-5, 
LH3, HH6, LH4, HH7, LH5, HH8, LH6, HH9, LH7, HH10, LH8, HH11, LH9, HH12, LH10, HH13, 
LH11, HH14, LH12, HH15, LH13, HH16, LH14, HH17, LH15 for $P_{sp} = -0.01\,C/m^2$; 
HH1-3, LH1, HH4, LH2, HH5-6, LH3, HH7, LH4, HH8, LH5, HH9-10, LH6, HH11, LH7, HH12, 
LH8, HH13, LH9, HH14, LH10, HH15-16, LH11, HH17, LH12-13, HH18, LH14 for 
$P_{sp} = -0.005\,C/m^2$ and HH1-7, LH1, HH8-9, LH2, HH10-11, LH3, HH12-13, 
LH4, HH14-15, LH5, HH16, LH6, HH17-18, LH7, HH19, LH8, HH20, LH9, HH21-22, LH10 
for $P_{sp} = -0.001\,C/m^2$. The number of HH states increases as the value of 
$P_{sp}$ decreases.

The anticrossings and also the spin splitting in the valence sub bands are more visible 
for higher values of $P_{sp}$. The strength of the resulting electric field not only 
increases the mixing of HH and LH states but also increases the value of the spin splitting 
in each sub band. This spin splitting is known as the Rashba effect \cite{jpc-17-6039} and is 
due to potential inversion asymmetry, even though the term $\alpha(\vec{\sigma}\times\vec{k})\cdot\vec{E}$ 
does not appear explicitly in the Hamiltonian \cite{book-winkler}.

The number of confined states decreases as the spontaneous polarization decreases. 
On the other hand, the energy difference between the conduction and valence band ground 
state increases as the spontaneous polarization decreases, blueshifting the interband energy 
transitions.

\begin{figure}[h]
\begin{center}
\includegraphics{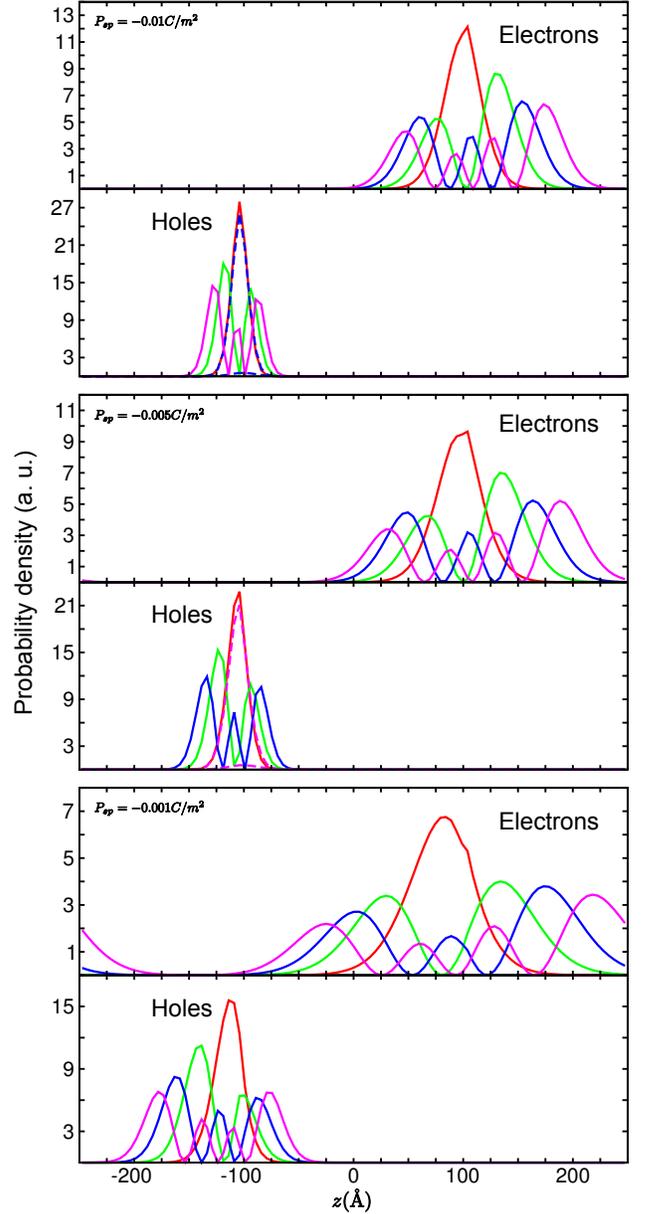}
\caption{Probability densities at $\vec{k}=0$ for the lowest four states of the conduction band and the 
highest four states of the valence band of Fig. \ref{fig:sppz_well_200_bs}.
The solid lines are the HH states and dashed lines represent the LH states.}
\label{fig:sppz_well_200_probdens}
\end{center}
\end{figure}

The effect of piezoelectric and spontaneous polarization also induce carriers' spatial 
separation. This effect, in the probability densities in $\vec{k}=0$ can be seen in Fig. 
\ref{fig:sppz_well_200_probdens}. The lowest four states of the conduction band and the 
highest four states of the valence band are presented. At $\vec{k}=0$ the wave functions of 
spin-up and spin-down are degenerated. We can see that the overlap increases for more excited 
states, also blueshifting the energy peak in the interband transitions. Since the potential 
profile is not completely even or odd, the envelope functions no longer have well defined parities. 

%================================================================================

\section{V. Conclusions}

The basic result of this study is the theoretical model based on the k$\cdot$p 
method and group theory concepts to calculate band structures of WZ/ZB polytypical 
systems in the vicinity of the band edge. The method allows us to describe in the 
same matrix Hamiltonian the ZB and WZ structures, with $k_z$ along the [111] and 
[0001] directions, respectively. Since the WZ structure is less symmetric, the ZB 
parameters are assigned to the WZ ones. Our method not only is able to describe 
the k$\cdot$p terms of the Hamiltonian but also includes the strain and polarization 
(spontaneous and piezoelectric) effects.

Extracting the parameters of WZ InP from Ref. \cite{prb-81-155210} we applied our 
model to a WZ/ZB/WZ single well in order to understand the physics of the polytypical 
interface. The potential profile at the interface WZ/ZB is type-II, whose feature is 
the spatial separation of carriers. The performed calculations in this study holds this 
characteristic.

Due to the lack of parameters in the literature for WZ InP, only the strain effect in 
the ZB region was considered here. This seems to be a reasonable consideration since 
the WZ structure is the dominant phase in NWs structures. However, such strain parameters 
would be fundamental in a system that the stable lattice constant is a intermediate 
value between WZ and ZB InP lattice parameters.

Within the limitation of strain, the piezoelectric polarization was also considered 
in the ZB region. For the WZ region, only the spontaneous polarization appears. Since 
there is no value in the literature for the spontaneous polarization of WZ InP, a 
range of values were considered in the simulations. Some of these values, however, 
induces a negative gap in the system. There is no data in the literature that corroborates 
this effect.

The proposed model, jointly with the obtained results, proved to be useful in the 
study of electronic band structures of WZ/ZB polytypical systems, such as NWs. 
Exploring the opportunities of band gap engineering considering not only different 
compounds, but also different crystal structures, could lead to the development of 
novel nanodevices.

%================================================================================

\section{Acknowledgements}

The authors acknowledge financial support from the Brazilian funding agencies 
CAPES and CNPq.

%================================================================================

\bibliography{referencias}

%Merlin.mbs v4.21 2009-07-09.
\begin{thebibliography}{10}%
\makeatletter
\providecommand \@ifxundefined [1]{%
 \ifx #1\undefined \expandafter \@firstoftwo
 \else \expandafter \@secondoftwo
\fi
}%
\providecommand \@ifnum [1]{%
 \ifnum #1\expandafter \@firstoftwo
 \else \expandafter \@secondoftwo
\fi
}%
\providecommand \enquote [1]{``#1''}%
\providecommand \bibnamefont  [1]{#1}%
\providecommand \bibfnamefont [1]{#1}%
\providecommand \citenamefont [1]{#1}%
\providecommand\href[0]{\@sanitize\@href}%
\providecommand\@href[1]{\endgroup\@@startlink{#1}\endgroup\@@href}%
\providecommand\@@href[1]{#1\@@endlink}%
\providecommand \@sanitize [0]{\begingroup\catcode`\&12\catcode`\#12\relax}%
\@ifxundefined \pdfoutput {\@firstoftwo}{%
 \@ifnum{\z@=\pdfoutput}{\@firstoftwo}{\@secondoftwo}%
}{%
 \providecommand\@@startlink[1]{\leavevmode\special{html:<a href="#1">}}%
 \providecommand\@@endlink[0]{\special{html:</a>}}%
}{%
 \providecommand\@@startlink[1]{%
  \leavevmode
  \pdfstartlink
   attr{/Border[0 0 1 ]/H/I/C[0 1 1]}%
   user{/Subtype/Link/A<</Type/Action/S/URI/URI(#1)>>}%
  \relax
 }%
 \providecommand\@@endlink[0]{\pdfendlink}%
}%
\providecommand \url  [0]{\begingroup\@sanitize \@url }%
\providecommand \@url [1]{\endgroup\@href {#1}{\urlprefix}}%
\providecommand \urlprefix [0]{URL }%
\providecommand \Eprint[0]{\href }%
\@ifxundefined \urlstyle {%
  \providecommand \doi [1]{doi:\discretionary{}{}{}#1}%
}{%
  \providecommand \doi [0]{doi:\discretionary{}{}{}\begingroup
  \urlstyle{rm}\Url }%
}%
\providecommand \doibase [0]{http://dx.doi.org/}%
\providecommand \Doi[1]{\href{\doibase#1}}%
\providecommand \bibAnnote [3]{%
  \BibitemShut{#1}%
  \begin{quotation}\noindent
    \textsc{Key:}\ #2\\\textsc{Annotation:}\ #3%
  \end{quotation}%
}%
\providecommand \bibAnnoteFile [2]{%
  \IfFileExists{#2}{\bibAnnote {#1} {#2} {\input{#2}}}{}%
}%
\providecommand \typeout [0]{\immediate \write \m@ne }%
\providecommand \selectlanguage [0]{\@gobble}%
\providecommand \bibinfo [0]{\@secondoftwo}%
\providecommand \bibfield [0]{\@secondoftwo}%
\providecommand \translation [1]{[#1]}%
\providecommand \BibitemOpen[0]{}%
\providecommand \bibitemStop [0]{}%
\providecommand \bibitemNoStop [0]{.\EOS\space}%
\providecommand \EOS [0]{\spacefactor3000\relax}%
\providecommand \BibitemShut [1]{\csname bibitem#1\endcsname}%
%</preamble>
\bibitem{science-293-1289}%
  \BibitemOpen
  \bibfield{author}{%
  \bibinfo {author} {\bibfnamefont{Y.}~\bibnamefont{Cui}}, \bibinfo {author}
  {\bibfnamefont{Q.}~\bibnamefont{Wei}}, \bibinfo {author}
  {\bibfnamefont{H.}~\bibnamefont{Park}},\ and\ \bibinfo {author}
  {\bibfnamefont{C.~M.}\ \bibnamefont{Lieber}},\ }%
  \bibfield{journal}{%
  \bibinfo {journal} {Science}\ }%
  \textbf{\bibinfo {volume} {293}},\ \bibinfo {pages} {1289} (\bibinfo {year}
  {2001})%
  \bibAnnoteFile{NoStop}{science-293-1289}%
\bibitem{nanomedicine-1-51}%
  \BibitemOpen
  \bibfield{author}{%
  \bibinfo {author} {\bibfnamefont{F.}~\bibnamefont{Patolsky}}, \bibinfo
  {author} {\bibfnamefont{G.}~\bibnamefont{Zheng}},\ and\ \bibinfo {author}
  {\bibfnamefont{L.~C.}\ \bibnamefont{M.}},\ }%
  \bibfield{journal}{%
  \bibinfo {journal} {Nanomedicine}\ }%
  \textbf{\bibinfo {volume} {1}},\ \bibinfo {pages} {51} (\bibinfo {year}
  {2006})%
  \bibAnnoteFile{NoStop}{nanomedicine-1-51}%
\bibitem{nbt-23-1294}%
  \BibitemOpen
  \bibfield{author}{%
  \bibinfo {author} {\bibfnamefont{G.}~\bibnamefont{Zheng}}, \bibinfo {author}
  {\bibfnamefont{F.}~\bibnamefont{Patolsky}}, \bibinfo {author}
  {\bibfnamefont{Y.}~\bibnamefont{Cui}}, \bibinfo {author}
  {\bibfnamefont{W.~U.}\ \bibnamefont{Wang}},\ and\ \bibinfo {author}
  {\bibfnamefont{C.~M.}\ \bibnamefont{Lieber}},\ }%
  \bibfield{journal}{%
  \bibinfo {journal} {Nature Biotechnology}\ }%
  \textbf{\bibinfo {volume} {23}},\ \bibinfo {pages} {1294} (\bibinfo {year}
  {2005})%
  \bibAnnoteFile{NoStop}{nbt-23-1294}%
\bibitem{nature-421-241}%
  \BibitemOpen
  \bibfield{author}{%
  \bibinfo {author} {\bibfnamefont{X.}~\bibnamefont{Duan}}, \bibinfo {author}
  {\bibfnamefont{Y.}~\bibnamefont{Huang}}, \bibinfo {author}
  {\bibfnamefont{R.}~\bibnamefont{Agarwal}},\ and\ \bibinfo {author}
  {\bibfnamefont{C.~M.}\ \bibnamefont{Lieber}},\ }%
  \bibfield{journal}{%
  \bibinfo {journal} {Nature}\ }%
  \textbf{\bibinfo {volume} {421}},\ \bibinfo {pages} {241} (\bibinfo {year}
  {2003})%
  \bibAnnoteFile{NoStop}{nature-421-241}%
\bibitem{nature-409-66}%
  \BibitemOpen
  \bibfield{author}{%
  \bibinfo {author} {\bibfnamefont{X.}~\bibnamefont{Duan}}, \bibinfo {author}
  {\bibfnamefont{Y.}~\bibnamefont{Huang}}, \bibinfo {author}
  {\bibfnamefont{Y.}~\bibnamefont{Cui}}, \bibinfo {author}
  {\bibfnamefont{J.}~\bibnamefont{Wang}},\ and\ \bibinfo {author}
  {\bibfnamefont{C.~M.}\ \bibnamefont{Lieber}},\ }%
  \bibfield{journal}{%
  \bibinfo {journal} {Nature}\ }%
  \textbf{\bibinfo {volume} {409}},\ \bibinfo {pages} {66} (\bibinfo {year}
  {2003})%
  \bibAnnoteFile{NoStop}{nature-409-66}%
\bibitem{nl-4-1247}%
  \BibitemOpen
  \bibfield{author}{%
  \bibinfo {author} {\bibfnamefont{H.}~\bibnamefont{Ng}}, \bibinfo {author}
  {\bibfnamefont{J.}~\bibnamefont{Han}}, \bibinfo {author}
  {\bibfnamefont{T.}~\bibnamefont{Yamada}},\ and\ \bibinfo {author}
  {\bibfnamefont{P.}~\bibnamefont{Nguyen}},\ }%
  \bibfield{journal}{%
  \bibinfo {journal} {Nano Letters}\ }%
  \textbf{\bibinfo {volume} {4}},\ \bibinfo {pages} {1247} (\bibinfo {year}
  {2004})%
  \bibAnnoteFile{NoStop}{nl-4-1247}%
\bibitem{apl-4-89}%
  \BibitemOpen
  \bibfield{author}{%
  \bibinfo {author} {\bibfnamefont{R.~S.}\ \bibnamefont{Wagner}}\ and\ \bibinfo
  {author} {\bibfnamefont{W.~C.}\ \bibnamefont{Ellis}},\ }%
  \bibfield{journal}{%
  \bibinfo {journal} {Applied Physics Letters}\ }%
  \textbf{\bibinfo {volume} {4}},\ \bibinfo {pages} {89} (\bibinfo {year}
  {1964})%
  \bibAnnoteFile{NoStop}{apl-4-89}%
\bibitem{semiconductors-43-1539}%
  \BibitemOpen
  \bibfield{author}{%
  \bibinfo {author} {\bibfnamefont{V.}~\bibnamefont{Dubrovskii}}, \bibinfo
  {author} {\bibfnamefont{G.}~\bibnamefont{Cirlin}},\ and\ \bibinfo {author}
  {\bibfnamefont{V.}~\bibnamefont{Ustinov}},\ }%
  \bibfield{journal}{%
  \bibinfo {journal} {Semiconductors}\ }%
  \textbf{\bibinfo {volume} {43}},\ \bibinfo {pages} {1539} (\bibinfo {year}
  {2009})%
  \bibAnnoteFile{NoStop}{semiconductors-43-1539}%
\bibitem{nl-10-1699}%
  \BibitemOpen
  \bibfield{author}{%
  \bibinfo {author} {\bibfnamefont{Y.}~\bibnamefont{Kitauchi}}, \bibinfo
  {author} {\bibfnamefont{Y.}~\bibnamefont{Kobayashi}}, \bibinfo {author}
  {\bibfnamefont{K.}~\bibnamefont{Tomioka}}, \bibinfo {author}
  {\bibfnamefont{S.}~\bibnamefont{Hara}}, \bibinfo {author}
  {\bibfnamefont{K.}~\bibnamefont{Hiruma}}, \bibinfo {author}
  {\bibfnamefont{T.}~\bibnamefont{Fukui}},\ and\ \bibinfo {author}
  {\bibfnamefont{J.}~\bibnamefont{Motohisa}},\ }%
  \bibfield{journal}{%
  \bibinfo {journal} {Nano Letters}\ }%
  \textbf{\bibinfo {volume} {10}},\ \bibinfo {pages} {1699} (\bibinfo {year}
  {2010})%
  \bibAnnoteFile{NoStop}{nl-10-1699}%
\bibitem{naturemat-5-574}%
  \BibitemOpen
  \bibfield{author}{%
  \bibinfo {author} {\bibfnamefont{J.}~\bibnamefont{Johansson}}, \bibinfo
  {author} {\bibfnamefont{L.~S.}\ \bibnamefont{Karlsson}}, \bibinfo {author}
  {\bibfnamefont{C.~P.~T.}\ \bibnamefont{Svensson}}, \bibinfo {author}
  {\bibfnamefont{T.}~\bibnamefont{Mårtensson}}, \bibinfo {author}
  {\bibfnamefont{B.~A.}\ \bibnamefont{Wacaser}}, \bibinfo {author}
  {\bibfnamefont{K.}~\bibnamefont{Deppert}}, \bibinfo {author}
  {\bibfnamefont{L.}~\bibnamefont{Samuelson}},\ and\ \bibinfo {author}
  {\bibfnamefont{W.}~\bibnamefont{Seifert}},\ }%
  \bibfield{journal}{%
  \bibinfo {journal} {Nature Materials}\ }%
  \textbf{\bibinfo {volume} {5}},\ \bibinfo {pages} {574} (\bibinfo {year}
  {2005})%
  \bibAnnoteFile{NoStop}{naturemat-5-574}%
\bibitem{naturenano-4-50}%
  \BibitemOpen
  \bibfield{author}{%
  \bibinfo {author} {\bibfnamefont{P.}~\bibnamefont{Caroff}}, \bibinfo {author}
  {\bibfnamefont{K.~A.}\ \bibnamefont{Dick}}, \bibinfo {author}
  {\bibfnamefont{J.}~\bibnamefont{Johansson}}, \bibinfo {author}
  {\bibfnamefont{M.~E.}\ \bibnamefont{Messing}}, \bibinfo {author}
  {\bibfnamefont{K.}~\bibnamefont{Deppert}},\ and\ \bibinfo {author}
  {\bibfnamefont{L.}~\bibnamefont{Samuelson}},\ }%
  \bibfield{journal}{%
  \bibinfo {journal} {Nature Nanotechnology}\ }%
  \textbf{\bibinfo {volume} {4}},\ \bibinfo {pages} {50} (\bibinfo {year}
  {2009})%
  \bibAnnoteFile{NoStop}{naturenano-4-50}%
\bibitem{nanoIEEE-6-384}%
  \BibitemOpen
  \bibfield{author}{%
  \bibinfo {author} {\bibfnamefont{S.-G.}\ \bibnamefont{Ihn}}, \bibinfo
  {author} {\bibfnamefont{J.-I.}\ \bibnamefont{Song}}, \bibinfo {author}
  {\bibfnamefont{Y.-H.}\ \bibnamefont{Kim}}, \bibinfo {author}
  {\bibfnamefont{J.~Y.}\ \bibnamefont{Lee}},\ and\ \bibinfo {author}
  {\bibfnamefont{I.-H.}\ \bibnamefont{Ahn}},\ }%
  \bibfield{journal}{%
  \bibinfo {journal} {IEEE Transactions on Nanotechnology}\ }%
  \textbf{\bibinfo {volume} {6}},\ \bibinfo {pages} {384} (\bibinfo {year}
  {2007})%
  \bibAnnoteFile{NoStop}{nanoIEEE-6-384}%
\bibitem{am-21-3654}%
  \BibitemOpen
  \bibfield{author}{%
  \bibinfo {author} {\bibfnamefont{J.}~\bibnamefont{Bao}}, \bibinfo {author}
  {\bibfnamefont{D.~C.}\ \bibnamefont{Bell}}, \bibinfo {author}
  {\bibfnamefont{F.}~\bibnamefont{Capasso}}, \bibinfo {author}
  {\bibfnamefont{N.}~\bibnamefont{Erdman}}, \bibinfo {author}
  {\bibfnamefont{D.}~\bibnamefont{Wei}}, \bibinfo {author}
  {\bibfnamefont{L.}~\bibnamefont{Fröberg}}, \bibinfo {author}
  {\bibfnamefont{T.}~\bibnamefont{Mårtensson}},\ and\ \bibinfo {author}
  {\bibfnamefont{L.}~\bibnamefont{Samuelson}},\ }%
  \bibfield{journal}{%
  \bibinfo {journal} {Advanced Materials}\ }%
  \textbf{\bibinfo {volume} {21}},\ \bibinfo {pages} {3654} (\bibinfo {year}
  {2009})%
  \bibAnnoteFile{NoStop}{am-21-3654}%
\bibitem{nano-22-265606}%
  \BibitemOpen
  \bibfield{author}{%
  \bibinfo {author} {\bibfnamefont{J.}~\bibnamefont{Bolinsson}}, \bibinfo
  {author} {\bibfnamefont{P.}~\bibnamefont{Caroff}}, \bibinfo {author}
  {\bibfnamefont{B.}~\bibnamefont{Mandl}},\ and\ \bibinfo {author}
  {\bibfnamefont{K.~A.}\ \bibnamefont{Dick}},\ }%
  \bibfield{journal}{%
  \bibinfo {journal} {Nanotechnology}\ }%
  \textbf{\bibinfo {volume} {22}},\ \bibinfo {pages} {265606} (\bibinfo {year}
  {2011})%
  \bibAnnoteFile{NoStop}{nano-22-265606}%
\bibitem{nl-11-2424}%
  \BibitemOpen
  \bibfield{author}{%
  \bibinfo {author} {\bibfnamefont{C.}~\bibnamefont{Thelander}}, \bibinfo
  {author} {\bibfnamefont{P.}~\bibnamefont{Caroff}}, \bibinfo {author}
  {\bibfnamefont{S.}~\bibnamefont{Plissard}}, \bibinfo {author}
  {\bibfnamefont{A.~W.}\ \bibnamefont{Dey}},\ and\ \bibinfo {author}
  {\bibfnamefont{K.~A.}\ \bibnamefont{Dick}},\ }%
  \bibfield{journal}{%
  \bibinfo {journal} {Nano Letters}\ }%
  \textbf{\bibinfo {volume} {11}},\ \bibinfo {pages} {2424} (\bibinfo {year}
  {2011})%
  \bibAnnoteFile{NoStop}{nl-11-2424}%
\bibitem{sst-25-024009}%
  \BibitemOpen
  \bibfield{author}{%
  \bibinfo {author} {\bibfnamefont{K.~A.}\ \bibnamefont{Dick}}, \bibinfo
  {author} {\bibfnamefont{P.}~\bibnamefont{Caroff}}, \bibinfo {author}
  {\bibfnamefont{J.}~\bibnamefont{Bolinsson}}, \bibinfo {author}
  {\bibfnamefont{M.~E.}\ \bibnamefont{Messing}}, \bibinfo {author}
  {\bibfnamefont{J.}~\bibnamefont{Johansson}}, \bibinfo {author}
  {\bibfnamefont{K.}~\bibnamefont{Deppert}}, \bibinfo {author}
  {\bibfnamefont{L.~R.}\ \bibnamefont{Wallenberg}},\ and\ \bibinfo {author}
  {\bibfnamefont{L.}~\bibnamefont{Samuelson}},\ }%
  \bibfield{journal}{%
  \bibinfo {journal} {Semiconductor Science and Technology}\ }%
  \textbf{\bibinfo {volume} {25}},\ \bibinfo {pages} {024009} (\bibinfo {year}
  {2010})%
  \bibAnnoteFile{NoStop}{sst-25-024009}%
\bibitem{pr-100-580}%
  \BibitemOpen
  \bibfield{author}{%
  \bibinfo {author} {\bibfnamefont{G.}~\bibnamefont{Dresselhaus}},\ }%
  \bibfield{journal}{%
  \bibinfo {journal} {Physical Review}\ }%
  \textbf{\bibinfo {volume} {100}},\ \bibinfo {pages} {580} (\bibinfo {year}
  {1955})%
  \bibAnnoteFile{NoStop}{pr-100-580}%
\bibitem{book-kane}%
  \BibitemOpen
  \bibfield{author}{%
  \bibinfo {author} {\bibfnamefont{E.~O.}\ \bibnamefont{Kane}},\ }%
  \emph{\bibinfo {title} {Physics of III-V compounds}},\ \bibinfo {series}
  {Semiconductors and Semimetals}, Vol.~\bibinfo {volume} {1}\ (\bibinfo
  {publisher} {Academic Press},\ \bibinfo {address} {New York},\ \bibinfo
  {year} {1966})%
  \bibAnnoteFile{NoStop}{book-kane}%
\bibitem{spjetp-14-898}%
  \BibitemOpen
  \bibfield{author}{%
  \bibinfo {author} {\bibfnamefont{G.~E.}\ \bibnamefont{Pikus}},\ }%
  \bibfield{journal}{%
  \bibinfo {journal} {Soviet Physics JETP}\ }%
  \textbf{\bibinfo {volume} {14}},\ \bibinfo {pages} {898} (\bibinfo {year}
  {1962})%
  \bibAnnoteFile{NoStop}{spjetp-14-898}%
\bibitem{book-birpikus}%
  \BibitemOpen
  \bibfield{author}{%
  \bibinfo {author} {\bibfnamefont{G.}~\bibnamefont{Bir}}\ and\ \bibinfo
  {author} {\bibfnamefont{G.}~\bibnamefont{Pikus}},\ }%
  \emph{\bibinfo {title} {Symmetry and strain-induced effects in
  semiconductors}}\ (\bibinfo {publisher} {Wiley},\ \bibinfo {address} {New
  York},\ \bibinfo {year} {1974})%
  \bibAnnoteFile{NoStop}{book-birpikus}%
\bibitem{prb-54-2491}%
  \BibitemOpen
  \bibfield{author}{%
  \bibinfo {author} {\bibfnamefont{S.~L.}\ \bibnamefont{Chuang}}\ and\ \bibinfo
  {author} {\bibfnamefont{C.~S.}\ \bibnamefont{Chang}},\ }%
  \bibfield{journal}{%
  \bibinfo {journal} {Physical Review B}\ }%
  \textbf{\bibinfo {volume} {54}},\ \bibinfo {pages} {2491} (\bibinfo {year}
  {1996})%
  \bibAnnoteFile{NoStop}{prb-54-2491}%
\bibitem{IEEEjqe-22-1625}%
  \BibitemOpen
  \bibfield{author}{%
  \bibinfo {author} {\bibfnamefont{G.}~\bibnamefont{Bastard}}\ and\ \bibinfo
  {author} {\bibfnamefont{J.~A.}\ \bibnamefont{Brum}},\ }%
  \bibfield{journal}{%
  \bibinfo {journal} {IEEE Journal of Quantum Electronics}\ }%
  \textbf{\bibinfo {volume} {22}},\ \bibinfo {pages} {1625} (\bibinfo {year}
  {1986})%
  \bibAnnoteFile{NoStop}{IEEEjqe-22-1625}%
\bibitem{prb-53-9930}%
  \BibitemOpen
  \bibfield{author}{%
  \bibinfo {author} {\bibfnamefont{G.~M.}\ \bibnamefont{Sipahi}}, \bibinfo
  {author} {\bibfnamefont{R.}~\bibnamefont{Enderlein}}, \bibinfo {author}
  {\bibfnamefont{L.~M.~R.}\ \bibnamefont{Scolfaro}},\ and\ \bibinfo {author}
  {\bibfnamefont{J.~R.}\ \bibnamefont{Leite}},\ }%
  \bibfield{journal}{%
  \bibinfo {journal} {Physical Review B}\ }%
  \textbf{\bibinfo {volume} {53}},\ \bibinfo {pages} {9930} (\bibinfo {year}
  {1996})%
  \bibAnnoteFile{NoStop}{prb-53-9930}%
\bibitem{apl-76-1015}%
  \BibitemOpen
  \bibfield{author}{%
  \bibinfo {author} {\bibfnamefont{S.~C.~P.}\ \bibnamefont{Rodrigues}},
  \bibinfo {author} {\bibfnamefont{G.~M.}\ \bibnamefont{Sipahi}}, \bibinfo
  {author} {\bibfnamefont{L.~M.~R.}\ \bibnamefont{Scolfaro}},\ and\ \bibinfo
  {author} {\bibfnamefont{J.~R.}\ \bibnamefont{Leite}},\ }%
  \bibfield{journal}{%
  \bibinfo {journal} {Applied Physics Letters}\ }%
  \textbf{\bibinfo {volume} {76}},\ \bibinfo {pages} {1015} (\bibinfo {year}
  {2000})%
  \bibAnnoteFile{NoStop}{apl-76-1015}%
\bibitem{sst-12-252}%
  \BibitemOpen
  \bibfield{author}{%
  \bibinfo {author} {\bibfnamefont{L.~C.}\ \bibnamefont{Chuang}}\ and\ \bibinfo
  {author} {\bibfnamefont{C.~S.}\ \bibnamefont{Chang}},\ }%
  \bibfield{journal}{%
  \bibinfo {journal} {Semiconductor Science and Technology}\ }%
  \textbf{\bibinfo {volume} {12}},\ \bibinfo {pages} {252} (\bibinfo {year}
  {1997})%
  \bibAnnoteFile{NoStop}{sst-12-252}%
\bibitem{jcg-246-347}%
  \BibitemOpen
  \bibfield{author}{%
  \bibinfo {author} {\bibfnamefont{S.~C.~P.}\ \bibnamefont{Rodrigues}}\ and\
  \bibinfo {author} {\bibfnamefont{G.~M.}\ \bibnamefont{Sipahi}},\ }%
  \bibfield{journal}{%
  \bibinfo {journal} {Journal of Crystal Growth}\ }%
  \textbf{\bibinfo {volume} {246}},\ \bibinfo {pages} {347} (\bibinfo {year}
  {2002})%
  \bibAnnoteFile{NoStop}{jcg-246-347}%
\bibitem{prb-49-4710}%
  \BibitemOpen
  \bibfield{author}{%
  \bibinfo {author} {\bibfnamefont{M.}~\bibnamefont{Murayama}}\ and\ \bibinfo
  {author} {\bibfnamefont{T.}~\bibnamefont{Nakayama}},\ }%
  \bibfield{journal}{%
  \bibinfo {journal} {Physical Review B}\ }%
  \textbf{\bibinfo {volume} {49}},\ \bibinfo {pages} {4710} (\bibinfo {year}
  {1994})%
  \bibAnnoteFile{NoStop}{prb-49-4710}%
\bibitem{prb-81-155210}%
  \BibitemOpen
  \bibfield{author}{%
  \bibinfo {author} {\bibfnamefont{A.}~\bibnamefont{De}}\ and\ \bibinfo
  {author} {\bibfnamefont{C.~E.}\ \bibnamefont{Pryor}},\ }%
  \bibfield{journal}{%
  \bibinfo {journal} {Physical Review B}\ }%
  \textbf{\bibinfo {volume} {81}},\ \bibinfo {pages} {155210} (\bibinfo {year}
  {2010})%
  \bibAnnoteFile{NoStop}{prb-81-155210}%
\bibitem{nl-9-648}%
  \BibitemOpen
  \bibfield{author}{%
  \bibinfo {author} {\bibfnamefont{K.}~\bibnamefont{Pemasiri}}, \bibinfo
  {author} {\bibfnamefont{M.}~\bibnamefont{Montazeri}}, \bibinfo {author}
  {\bibfnamefont{R.}~\bibnamefont{Gass}}, \bibinfo {author}
  {\bibfnamefont{L.~M.}\ \bibnamefont{Smith}}, \bibinfo {author}
  {\bibfnamefont{H.~E.}\ \bibnamefont{Jackson}}, \bibinfo {author}
  {\bibfnamefont{J.}~\bibnamefont{Yarrison-Rice}}, \bibinfo {author}
  {\bibfnamefont{S.}~\bibnamefont{Paiman}}, \bibinfo {author}
  {\bibfnamefont{Q.}~\bibnamefont{Gao}}, \bibinfo {author}
  {\bibfnamefont{H.~H.}\ \bibnamefont{Tan}}, \bibinfo {author}
  {\bibfnamefont{C.}~\bibnamefont{Jagadish}}, \bibinfo {author}
  {\bibfnamefont{X.}~\bibnamefont{Zhang}},\ and\ \bibinfo {author}
  {\bibfnamefont{J.}~\bibnamefont{Zou}},\ }%
  \bibfield{journal}{%
  \bibinfo {journal} {Nano Letters}\ }%
  \textbf{\bibinfo {volume} {9}},\ \bibinfo {pages} {648} (\bibinfo {year}
  {2009})%
  \bibAnnoteFile{NoStop}{nl-9-648}%
\bibitem{nano-20-225606}%
  \BibitemOpen
  \bibfield{author}{%
  \bibinfo {author} {\bibfnamefont{S.}~\bibnamefont{Paiman}}, \bibinfo {author}
  {\bibfnamefont{Q.}~\bibnamefont{Gao}}, \bibinfo {author}
  {\bibfnamefont{H.~H.}\ \bibnamefont{Tan}}, \bibinfo {author}
  {\bibfnamefont{C.}~\bibnamefont{Jagadish}}, \bibinfo {author}
  {\bibfnamefont{K.}~\bibnamefont{Pemasiri}}, \bibinfo {author}
  {\bibfnamefont{M.}~\bibnamefont{Montazeri}}, \bibinfo {author}
  {\bibfnamefont{H.~E.}\ \bibnamefont{Jackson}}, \bibinfo {author}
  {\bibfnamefont{L.~M.}\ \bibnamefont{Smith}}, \bibinfo {author}
  {\bibfnamefont{J.~M.}\ \bibnamefont{Yarrison-Rice}}, \bibinfo {author}
  {\bibfnamefont{X.}~\bibnamefont{Zhang}},\ and\ \bibinfo {author}
  {\bibfnamefont{J.}~\bibnamefont{Zou}},\ }%
  \bibfield{journal}{%
  \bibinfo {journal} {Nanotechnology}\ }%
  \textbf{\bibinfo {volume} {20}},\ \bibinfo {pages} {225606} (\bibinfo {year}
  {2009})%
  \bibAnnoteFile{NoStop}{nano-20-225606}%
\bibitem{prb-82-125327}%
  \BibitemOpen
  \bibfield{author}{%
  \bibinfo {author} {\bibfnamefont{E.~G.}\ \bibnamefont{Gadret}}, \bibinfo
  {author} {\bibfnamefont{G.~O.}\ \bibnamefont{Dias}}, \bibinfo {author}
  {\bibfnamefont{L.~C.~O.}\ \bibnamefont{Dacal}}, \bibinfo {author}
  {\bibfnamefont{M.~M.}\ \bibnamefont{de~Lima~Jr.}}, \bibinfo {author}
  {\bibfnamefont{C.~V. R.~S.}\ \bibnamefont{Ruffo}}, \bibinfo {author}
  {\bibfnamefont{F.}~\bibnamefont{Iikawa}}, \bibinfo {author}
  {\bibfnamefont{M.~J. S.~P.}\ \bibnamefont{Brasil}}, \bibinfo {author}
  {\bibfnamefont{T.}~\bibnamefont{Chiaramonte}}, \bibinfo {author}
  {\bibfnamefont{M.~A.}\ \bibnamefont{Cotta}}, \bibinfo {author}
  {\bibfnamefont{L.~H.~G.}\ \bibnamefont{Tizei}}, \bibinfo {author}
  {\bibfnamefont{D.}~\bibnamefont{Ugarte}},\ and\ \bibinfo {author}
  {\bibfnamefont{A.}~\bibnamefont{Cantarero}},\ }%
  \bibfield{journal}{%
  \bibinfo {journal} {Physical Review B}\ }%
  \textbf{\bibinfo {volume} {82}},\ \bibinfo {pages} {125327} (\bibinfo {year}
  {2010})%
  \bibAnnoteFile{NoStop}{prb-82-125327}%
\bibitem{nanolet-10-4055}%
  \BibitemOpen
  \bibfield{author}{%
  \bibinfo {author} {\bibfnamefont{L.}~\bibnamefont{Zhang}}, \bibinfo {author}
  {\bibfnamefont{J.-W.}\ \bibnamefont{Luo}}, \bibinfo {author}
  {\bibfnamefont{A.}~\bibnamefont{Zunger}}, \bibinfo {author}
  {\bibfnamefont{N.}~\bibnamefont{Akopian}}, \bibinfo {author}
  {\bibfnamefont{V.}~\bibnamefont{Zwiller}},\ and\ \bibinfo {author}
  {\bibfnamefont{J.-C.}\ \bibnamefont{Harmand}},\ }%
  \bibfield{journal}{%
  \bibinfo {journal} {Nanoletters}\ }%
  \textbf{\bibinfo {volume} {10}},\ \bibinfo {pages} {4055} (\bibinfo {year}
  {2010})%
  \bibAnnoteFile{NoStop}{nanolet-10-4055}%
\bibitem{nanotec-21-505709}%
  \BibitemOpen
  \bibfield{author}{%
  \bibinfo {author} {\bibfnamefont{D.}~\bibnamefont{Li}}, \bibinfo {author}
  {\bibfnamefont{Z.}~\bibnamefont{Wang}},\ and\ \bibinfo {author}
  {\bibfnamefont{F.}~\bibnamefont{Gao}},\ }%
  \bibfield{journal}{%
  \bibinfo {journal} {Nanotechnology}\ }%
  \textbf{\bibinfo {volume} {21}},\ \bibinfo {pages} {505709} (\bibinfo {year}
  {2010})%
  \bibAnnoteFile{NoStop}{nanotec-21-505709}%
\bibitem{ssc-151-781}%
  \BibitemOpen
  \bibfield{author}{%
  \bibinfo {author} {\bibfnamefont{L.~C.~O.}\ \bibnamefont{Dacal}}\ and\
  \bibinfo {author} {\bibfnamefont{A.}~\bibnamefont{Cantarero}},\ }%
  \bibfield{journal}{%
  \bibinfo {journal} {Solid State Communications}\ }%
  \textbf{\bibinfo {volume} {151}},\ \bibinfo {pages} {781} (\bibinfo {year}
  {2011})%
  \bibAnnoteFile{NoStop}{ssc-151-781}%
\bibitem{jap-104-044313}%
  \BibitemOpen
  \bibfield{author}{%
  \bibinfo {author} {\bibfnamefont{M.}~\bibnamefont{Moewe}}, \bibinfo {author}
  {\bibfnamefont{L.~C.}\ \bibnamefont{Chuang}}, \bibinfo {author}
  {\bibfnamefont{V.~G.}\ \bibnamefont{Dubrovskii}},\ and\ \bibinfo {author}
  {\bibfnamefont{C.}~\bibnamefont{Chang-Hasnain}},\ }%
  \bibfield{journal}{%
  \bibinfo {journal} {Journal of Applied Physics}\ }%
  \textbf{\bibinfo {volume} {104}},\ \bibinfo {pages} {044313} (\bibinfo {year}
  {2008})%
  \bibAnnoteFile{NoStop}{jap-104-044313}%
\bibitem{dresselhaus-jorio}%
  \BibitemOpen
  \bibfield{author}{%
  \bibinfo {author} {\bibfnamefont{M.~S.}\ \bibnamefont{Dresselhaus}}, \bibinfo
  {author} {\bibfnamefont{G.}~\bibnamefont{Dresselhaus}},\ and\ \bibinfo
  {author} {\bibfnamefont{A.}~\bibnamefont{Jorio}},\ }%
  \emph{\bibinfo {title} {Group Theory: application to the physics of condensed
  matter}}\ (\bibinfo {publisher} {Springer-Verlag},\ \bibinfo {year} {2008})%
  \bibAnnoteFile{NoStop}{dresselhaus-jorio}%
\bibitem{arxiv}%
  \BibitemOpen
  \bibfield{author}{%
  \bibinfo {author} {\bibfnamefont{P.~E.}\ \bibnamefont{Faria~Junior}}\ and\
  \bibinfo {author} {\bibfnamefont{G.~M.}\ \bibnamefont{Sipahi}},\ }%
  \bibfield{journal}{%
  \bibinfo {journal} {http://arxiv.org/abs/1012.0227}}%
   (\bibinfo {year} {2010})%
  \bibAnnoteFile{NoStop}{arxiv}%
\bibitem{jap-87-353}%
  \BibitemOpen
  \bibfield{author}{%
  \bibinfo {author} {\bibfnamefont{S.~H.}\ \bibnamefont{Park}}\ and\ \bibinfo
  {author} {\bibfnamefont{S.~L.}\ \bibnamefont{Chuang}},\ }%
  \bibfield{journal}{%
  \bibinfo {journal} {Journal of Applied Physics}\ }%
  \textbf{\bibinfo {volume} {87}},\ \bibinfo {pages} {353} (\bibinfo {year}
  {2000})%
  \bibAnnoteFile{NoStop}{jap-87-353}%
\bibitem{jap-89-5815}%
  \BibitemOpen
  \bibfield{author}{%
  \bibinfo {author} {\bibfnamefont{I.}~\bibnamefont{Vurgaftman}}, \bibinfo
  {author} {\bibfnamefont{J.~R.}\ \bibnamefont{Meyer}},\ and\ \bibinfo {author}
  {\bibfnamefont{L.~R.}\ \bibnamefont{Ram-Mohan}},\ }%
  \bibfield{journal}{%
  \bibinfo {journal} {Journal of Applied Physics}\ }%
  \textbf{\bibinfo {volume} {89}},\ \bibinfo {pages} {5815} (\bibinfo {year}
  {2001})%
  \bibAnnoteFile{NoStop}{jap-89-5815}%
\bibitem{apl-68-1657}%
  \BibitemOpen
  \bibfield{author}{%
  \bibinfo {author} {\bibfnamefont{S.~L.}\ \bibnamefont{Chuang}}\ and\ \bibinfo
  {author} {\bibfnamefont{C.~S.}\ \bibnamefont{Chang}},\ }%
  \bibfield{journal}{%
  \bibinfo {journal} {Applied Physics Letters}\ }%
  \textbf{\bibinfo {volume} {68}},\ \bibinfo {pages} {1657} (\bibinfo {year}
  {1996})%
  \bibAnnoteFile{NoStop}{apl-68-1657}%
\bibitem{prb-35-1242}%
  \BibitemOpen
  \bibfield{author}{%
  \bibinfo {author} {\bibfnamefont{C.}~\bibnamefont{Mailhiot}}\ and\ \bibinfo
  {author} {\bibfnamefont{D.~L.}\ \bibnamefont{Smith}},\ }%
  \bibfield{journal}{%
  \bibinfo {journal} {Physical Review B}\ }%
  \textbf{\bibinfo {volume} {35}},\ \bibinfo {pages} {1242} (\bibinfo {year}
  {1987})%
  \bibAnnoteFile{NoStop}{prb-35-1242}%
\bibitem{paul-harrison}%
  \BibitemOpen
  \bibfield{author}{%
  \bibinfo {author} {\bibfnamefont{P.}~\bibnamefont{Harrison}},\ }%
  \emph{\bibinfo {title} {Quantum Wells, Wires and Dots: Theoretical and
  Computational Physics of Semiconductor Nanostructures}}\ (\bibinfo
  {publisher} {John Wiley and Sons Ltd},\ \bibinfo {year} {2006})%
  \bibAnnoteFile{NoStop}{paul-harrison}%
\bibitem{book-bastard}%
  \BibitemOpen
  \bibfield{author}{%
  \bibinfo {author} {\bibfnamefont{G.}~\bibnamefont{Bastard}},\ }%
  \emph{\bibinfo {title} {Wave mechanics applied to semiconductor
  heterostructures}}\ (\bibinfo {publisher} {Halsted Press},\ \bibinfo
  {address} {Les Ulis Cedex, France},\ \bibinfo {year} {1988})%
  \bibAnnoteFile{NoStop}{book-bastard}%
\bibitem{jap-92-932}%
  \BibitemOpen
  \bibfield{author}{%
  \bibinfo {author} {\bibfnamefont{H.}~\bibnamefont{Magnus}}, \bibinfo {author}
  {\bibfnamefont{M.~E.}\ \bibnamefont{Pistol}},\ and\ \bibinfo {author}
  {\bibfnamefont{C.}~\bibnamefont{Pryor}},\ }%
  \bibfield{journal}{%
  \bibinfo {journal} {Journal of Applied Physics}\ }%
  \textbf{\bibinfo {volume} {92}},\ \bibinfo {pages} {932} (\bibinfo {year}
  {2002})%
  \bibAnnoteFile{NoStop}{jap-92-932}%
\bibitem{jpc-17-6039}%
  \BibitemOpen
  \bibfield{author}{%
  \bibinfo {author} {\bibfnamefont{Y.~A.}\ \bibnamefont{Bychkov}}\ and\
  \bibinfo {author} {\bibfnamefont{E.~I.}\ \bibnamefont{Rashba}},\ }%
  \bibfield{journal}{%
  \bibinfo {journal} {Journal of Physics C: Solid State Physics}\ }%
  \textbf{\bibinfo {volume} {17}},\ \bibinfo {pages} {6039} (\bibinfo {year}
  {1984})%
  \bibAnnoteFile{NoStop}{jpc-17-6039}%
\bibitem{book-winkler}%
  \BibitemOpen
  \bibfield{author}{%
  \bibinfo {author} {\bibfnamefont{R.}~\bibnamefont{Winkler}},\ }%
  \emph{\bibinfo {title} {Spin-orbit coupling effects in two-dimensional
  electron and hole systems}}\ (\bibinfo {publisher} {Springer},\ \bibinfo
  {address} {Berlin},\ \bibinfo {year} {2003})%
  \bibAnnoteFile{NoStop}{book-winkler}%
\end{thebibliography}%

\end{document}